\documentclass[a4paper,fleqn,usenatbib,useAMS]{mnras}

\usepackage{graphicx}   % Including figure files
\usepackage{amsmath}    % Advanced maths commands
\usepackage{amssymb}    % Extra maths symbols
\usepackage{multicol}        % Multi-column entries in tables
\usepackage{bm}     % Bold maths symbols, including upright Greek
\usepackage{pdflscape}  % Landscape pages
\usepackage[caption = false]{subfig}
\usepackage{lipsum}
\usepackage[T1]{fontenc}
\usepackage{ae,aecompl}

\usepackage{txfonts}

\title[The impact of baryons on massive galaxy clusters]{The impact of baryons on massive galaxy clusters: halo structure and cluster mass estimates}
\author[M. A. Henson]{Monique A. Henson$^1$\thanks{Contact email: monique.henson@manchester.ac.uk}, David J. Barnes$^1$, Scott T. Kay$^1$, Ian G. McCarthy$^2$ and Joop Schaye$^3$ \\
$^1$Jodrell Bank Centre for Astrophysics, School of Physics and Astronomy, The University of Manchester, Manchester M13 9PL \\
$^2$Astrophysics Research Institute, Liverpool John Moores University, 146 Brownlow Hill, Liverpool L3 5RF \\
$^3$Leiden Observatory, Leiden University, P.O. Box 9513, 2300 RA Leiden, the Netherlands}

\pubyear{2016}

\begin{document}
\pagerange{\pageref{firstpage}--\pageref{lastpage}}
\label{firstpage}
\maketitle

\begin{abstract}
We use the BAHAMAS and MACSIS hydrodynamic simulations to quantify the impact of baryons on the mass distribution and dynamics of massive galaxy clusters, as well as the bias in X-ray and weak lensing mass estimates. These simulations use the sub-grid physics models calibrated in the BAHAMAS project, which include feedback from both supernovae and active galactic nuclei. They form a cluster population covering almost two orders of magnitude in mass, with more than 3,500 clusters with masses greater than $10^{14}\,\mathrm{M}_\odot$ at $z=0$. We start by characterising the clusters in terms of their spin, shape and density profile, before considering the bias in both weak lensing and hydrostatic mass estimates. Whilst including baryonic effects leads to more spherical, centrally concentrated clusters, the median weak lensing mass bias is unaffected by the presence of baryons. In both the dark matter only and hydrodynamic simulations, the weak lensing measurements underestimate cluster masses by ${\approx}10\%$ for clusters with $M_{200}{\leq}10^{15}\mathrm{M}_\odot$ and this bias tends to zero at higher masses. We also consider the hydrostatic bias when using both the true density and temperature profiles, and those derived from X-ray spectroscopy. When using spectroscopic temperatures and densities, the hydrostatic bias decreases as a function of mass, leading to a bias of ${\approx}40\%$ for clusters with $M_{500}{\geq}10^{15}\,\mathrm{M}_\odot$. This is due to the presence of cooler gas in the cluster outskirts. Using mass weighted temperatures and the true density profile reduces this bias to $5{-}15\%$. 
\end{abstract}


\section{Introduction}
Galaxy clusters are a sensitive probe of the late time evolution of the Universe, providing crucial insights into the nature of both dark matter and dark energy. Cluster based cosmological tests require well constrained masses for large samples of clusters. There is a long standing debate about the bias in X-ray cluster masses \citep[see][]{Mazzotta2004,Rasia2012,Applegate2014,Smith2016}, which arises due to the assumption that clusters are in hydrostatic equilibrium. Since clusters are often unrelaxed systems, this is frequently not a valid assumption. Instead, many authors are moving towards using masses derived from weak lensing (WL) observations of clusters~\citep{Okabe2010,Mahdavi2013,Hoekstra2015,Kettula2015} or at the very least, calibrating X-ray masses using weak lensing measurements~\citep[e.g.][]{Lieu2015}. The power of cluster counting has been highlighted in Sunyaev-Zel'dovich surveys performed by the \textit{Planck} satellite~\citep{Planck2015} and the South Pole Telescope~\citep{Bocquet2015}, however more accurate cluster mass measurements are needed for cluster cosmology to be competitive with other techniques~\citep{Allen2011,Planck2015}. High quality observational data is forthcoming with the ongoing and upcoming Dark Energy Survey~\citep{DES2005}, SPT-3G~\citep{Benson2014}, Large Synoptic Sky Survey~\citep{Ivezic2008} and ACTpol~\citep{Niemack2010}, but we also need simulations to provide robust theoretical predictions for comparison, as well mock data for testing observational techniques. 

Galaxy clusters have been extensively studied in dark matter only (DMO) simulations. It is well established in those simulations that cold dark matter haloes are triaxial, prolate structures. The sphericity of dark matter haloes decreases with increasing mass, so that galaxy clusters typically have sphericities of $(c/a){\simeq}0.4{-}0.6$~\citep{Maccio2008,Muonoz-Cuartas2011,Bryan2013}. Since both concentration and spin have also been shown to decrease weakly with mass~\citep{Bett2007,Duffy2008,Klypin2011,Muonoz-Cuartas2011,Ludlow2012,Klypin2016}, high mass clusters typically have low concentrations and exhibit little rotational support. 

DMO simulations have also been instrumental in testing observational methods for measuring cluster masses. Weak gravitational lensing provides a promising method for measuring the masses of galaxy clusters, since it does not require any assumptions about the dynamical state of the cluster. DMO simulations have shown that weak lensing masses are typically biased low by ${\sim}5\%$, with this bias decreasing with increasing mass~\citep{Oguri2011,Becker2011,Bahe2012}. Understanding this bias is crucial for cluster cosmology, since it requires large samples of clusters with accurately determined masses. 

Cosmological hydrodynamic simulations have shown that including baryons can have a significant effect upon the mass distribution of groups and low-mass clusters~\citep[e.g.][]{Bryan2013,Velliscig2014,Cusworth2014,Schaller2015}. The inclusion of baryonic effects in cosmological simulations leads to the depletion of high mass clusters~\citep{Cusworth2014}, and the clusters that do form are more spherical and have higher concentrations than their dark matter only counterparts~\citep{Duffy2010,Bryan2013}. The baryon fraction and hence the total mass within clusters is sensitive to galaxy formation processes~\citep{Stanek2009,McCarthy2011,Martizzi2012,Velliscig2014,LeBrun2014}. 

Thus, the impact of baryons on the shape and density profile of clusters depends on galaxy formation efficiency~\citep{Bryan2013,Duffy2010}. The impact of baryons on the mass distribution of low-mass clusters is not just limited to the central regions of clusters; feedback from Active Galactic Nuclei (AGN) can alter low-mass cluster profiles out to $R_{200}$~\citep{Velliscig2014}\footnote{$M_\Delta$ is defined as the mass contained within a sphere of radius $R_\Delta$, at which the enclosed average density is $\Delta$ times the critical density of the Universe}. It is still unclear what effect baryons will have on high-mass clusters. If baryons have a significant impact on the mass distribution of massive galaxy clusters, this may have implications for mass estimation techniques such as cluster weak lensing, which have been tested on dark matter only simulations~\citep{Oguri2011,Becker2011,Bahe2012}.

The lack of hydrodynamic simulations of massive galaxy clusters is a natural consequence of the large computational cost of such simulations. Furthermore, accounting for baryonic effects is not a trivial task, requiring calibrated models for star formation, feedback from supernovae and AGN, and radiative cooling. Cosmological zoom simulations, in which the region of interest in simulated at a higher resolution than the surrounding region, offer a solution to this problem. This approach has been used on cluster scales \citep[e.g.][]{Martizzi2014a,Hahn2015}, however it has only been applied to small numbers of clusters to date. This places limitations on the conclusions of such work, since the dynamic range in mass needed to investigate mass dependent properties is lacking and it is difficult to determine whether any results are significant or an artefact of the small sample size.

To obtain a sample sufficiently large to investigate the properties of massive galaxy clusters, we combine the $400\,h^{-1}\mathrm{Mpc}$ BAryons and HAloes of MAssive Systems (BAHAMAS) simulation~\citep{McCarthy2016} with the hydrodynamic zoom simulations that were developed as part of the MAssive ClusterS and Intercluster Structures (MACSIS) project~\citep{Barnes2016}.

The paper is organised as follows. The simulations used and the methods used to identify haloes and classify relaxed structures are described in Section 2. In Section 3 the methods used to measure the spins, shapes and density profiles of clusters are outlined and results are presented. This is followed by the results from a weak lensing analysis of the cluster sample in Section 4. In Section 5 we discuss hydrostatic bias in this cluster sample and the method used to calculate the X-ray hydrostatic masses. Finally, we summarise our results in Section 6.

\section{Simulations}

\subsection{BAHAMAS \label{sec:BAHAMAS}}
The BAHAMAS simulations relevant to this work consist of a dark matter only simulation (hereafter BAHAMAS-DMO) and a baryonic simulation (hereafter BAHAMAS-HYDRO), which consist of $2{\times} 1024^3$ particles in boxes with sides of length 400\,$h^{-1}$ (comoving) Mpc in the \textit{Planck} cosmology \citep{Planck2015}. The key cosmological parameters are given in Table~\ref{tbl:cosmological-parameters}. For the BAHAMAS-HYDRO simulations, the smoothed particle hydrodynamics code GADGET-3~\citep{Springel2005} has been modified to incorporate sub-grid prescriptions developed for the OWLS project~\citep{Schaye2010}, which model the effects of radiative cooling~\citep{Wiersma2009}, star formation~\citep{Schaye2008} and feedback from AGN~\citep{Booth2009} and supernovae~\citep{DallaVecchia2008}. The calibration of the models for stellar and AGN feedback is described in~\cite{McCarthy2016}. Briefly, the feedback models (both AGN and supernovae) were calibrated to reproduce the observed gas fractions of groups and clusters~\citep{Vikhlinin2006,Maughan2008,Sun2009,Pratt2009,Lin2012} and the global galaxy stellar mass function~\citep{Bernardi2013,Baldry2012,Li2009}.

As shown in~\cite{McCarthy2016}, the BAHAMAS simulations reproduce both the observed stellar and hot gas properties of groups and clusters, including the observed stellar mass fractions of central galaxies, and the amplitude of the relation between the integrated stellar mass fraction and halo mass for groups and clusters. BAHAMAS also recovers the observed X-ray and Sunyaev-Zel'dovich scaling relations, in addition to their observed pressure and density profiles. 

\subsection{MACSIS}
\renewcommand{\tabcolsep}{4.5pt}
\begin{table}
\centering
\caption{Cosmological parameters used in the BAHAMAS and MACSIS simulations. All values are consistent with~\protect\cite{Planck2015}.}
\begin{tabular}{p{1.4cm}|c|c|c|c|c|c|c}%p{0.6cm}p{.6cm}p{.6cm}p{.6cm}p{.6cm}p{.6cm}}
\hline
Simulation(s) & $\Omega_\mathrm{m}$ & $\Omega_\mathrm{b} $ & $\Omega_\Lambda$ & $\sigma_8$ & $n_s$ & $h$ \\
\hline
BAHAMAS & 0.3175 & 0.04900 & 0.6825 &  0.8340 & 0.9624 & 0.6711 \\
MACSIS & 0.3070 & 0.04825 & 0.6930 & 0.8288 & 0.9611 & 0.6777 \\
\hline
\end{tabular}
\label{tbl:cosmological-parameters}
\end{table}
\renewcommand{\tabcolsep}{6pt}
The MACSIS project is a set of cosmological simulations of massive galaxy clusters described in depth in~\cite{Barnes2016}.

The foundation of the project is a 3.2\,Gpc DMO simulation (hereafter referred to as the ``parent'' simulation), which adopts the \textit{Planck} cosmology~\citep{Planck2015}. The large spatial extent of this parent simulation allows for the inclusion of longer wavelength perturbations in the initial conditions, which leads to the formation of rarer, more massive structures. At $z=0$, the parent simulation contains more than 100,000 haloes with $M_{200}{\geq}10^{14}\,h^{-1}\mathrm{M}_\odot$.  This simulation has a softening length of $40\,h^{-1}$kpc at $z=0$ and a dark matter particle mass of $5.43{\times}10^{10}\,h^{-1}\mathrm{M}_\odot$.

A sample of 390 haloes from this parent box were selected for resimulation at higher resolution with the BAHAMAS model. Haloes in the parent simulation were binned by Friends-of-Friends (FoF) mass in bins of width $\Delta\log_{10}M_\mathrm{FoF}=0.2$ between $10^{15}{\leq} M_\mathrm{FoF}/\mathrm{M}_\odot{\leq}10^{16}$. Below $M_\mathrm{FoF}=10^{15.6}\,\mathrm{M}_\odot$, each of these bins was further divided into ten bins, within which 10 haloes were selected at random to produce a sample of 300 haloes. We have verified that the spins, shapes and concentrations of these haloes are consistent with the underlying parent population. In the parent simulation there are 90 haloes with masses $M_\mathrm{FoF}{\geq}10^{15.6}\,\mathrm{M}_\odot$. The most massive halo in the parent box has a FoF mass of $M_\mathrm{FoF}=10^{15.8}\,h^{-1}\mathrm{M}_\odot$. All of the most massive 90 haloes were selected for resimulation, producing an overall sample of 390 haloes.

The region around each cluster was resimulated  at a higher resolution using the OWLS version of GADGET-3. The resolution of the initial conditions of the parent simulation were progressively degraded with increasing distance from the high resolution region. This approach includes the large scale power and tidal forces from the parent box, whilst achieving the desired resolution in the region surrounding the cluster. 

Two resimulations were performed for each cluster: one dark matter only simulation (MACSIS-DMO) with a particle mass of $5.2{\times}10^9\,h^{-1}\mathrm{M}_\odot$, and a hydrodynamical simulation (MACSIS-HYDRO). 
The hydrodynamical simulations used the BAHAMAS code detailed in Section~\ref{sec:BAHAMAS}, with a dark matter particle mass of $4.4{\times} 10^9\,h^{-1}\mathrm{M}_\odot$ and an initial gas particle mass of $8.0{\times} 10^8\,h^{-1}\mathrm{M}_\odot$. In the MACSIS and BAHAMAS simulations considered here, the gravitational softening length was set to $4\,h^{-1}\mathrm{kpc}$ in physical coordinates for $z{\leq}3$ and $16\,h^{-1}\mathrm{Mpc}$ in comoving coordinates for $z{>}3$.

\begin{table*} 
\centering 
\caption{Mass cuts made to the BAHAMAS (BAH) and MACSIS (MAC) simulations at various redshifts and the number of clusters above various minimum mass limits. Only BAHAMAS clusters with $M_{200}{\leq} M_\mathrm{cut}$ and MACSIS clusters with $M_{200}{\geq}M_\mathrm{cut}$ are included in the cluster sample. $N_\mathrm{cut}$ is the number of haloes removed in the mass cuts in the MACSIS and BAHAMAS simulations. The outputs of the BAHAMAS and MACSIS simulations at $z{\approx}0.25,0.5$ are at slightly different redshifts. As a consequence, they are not used when looking at any property that may be redshift dependent.}
\begin{tabular}{c|c|c|c|c|c|c|c|c|c|c|c}
\hline 
\multicolumn{2}{c}{$z$} & 	$M_\mathrm{cut}/h^{-1}\mathrm{M}_\odot$ & 	\multicolumn{2}{c}{$N_\mathrm{cut}$} & 	\multicolumn{2}{c}{$N(M_{200}{\geq}5{\times}10^{13}h^{-1}\mathrm{M}_\odot)$} & 	\multicolumn{2}{c}{$N(M_{200}{\geq}1{\times}10^{14}h^{-1}\mathrm{M}_\odot)$} & 	\multicolumn{2}{c}{$N(M_{500}{\geq}1{\times}10^{14}h^{-1}\mathrm{M}_\odot)$}  	\\
MAC & BAH & & MAC & BAH & MAC & BAH & MAC & BAH & MAC & BAH \\[2pt] 
\hline 
HYDRO \\[2pt] 
$0.00$ & 	$0.00$ & 	$10^{14.7}$ & 	$~59$ & 	$~48$ & 	$~331$ & 	$3250$ & 	$~331$ & 	$1192$ & 	$~331$ & 	$~637$  	\\[2pt] 
$0.24$ & 	$0.25$ & 	$10^{14.5}$ & 	$~46$ & 	$~81$ & 	$~344$ & 	$2668$ & 	$~344$ & 	$~858$ & 	$~344$ & 	$~397$  	\\[2pt] 
$0.46$ & 	$0.50$ & 	$10^{14.3}$ & 	$~33$ & 	$124$ & 	$~357$ & 	$1956$ & 	$~357$ & 	$~521$ & 	$~355$ & 	$~143$  	\\[2pt] 
$1.00$ & 	$1.00$ & 	$10^{14.1}$ & 	$~90$ & 	$~86$ & 	$~300$ & 	$~766$ & 	$~300$ & 	$~~91$ & 	$~252$ & 	$~~~1$  	\\[2pt] 
 \\[2pt] 
DMO   \\[2pt] 
$0.00$ & 	$0.00$ & 	$10^{14.7}$ & 	$~58$ & 	$~48$ & 	$~332$ & 	$3553$ & 	$~332$ & 	$1267$ & 	$~332$ & 	$~675$  	\\[2pt] 
$0.24$ & 	$0.25$ & 	$10^{14.5}$ & 	$~44$ & 	$~72$ & 	$~346$ & 	$2917$ & 	$~346$ & 	$~923$ & 	$~346$ & 	$~436$  	\\[2pt] 
$0.46$ & 	$0.50$ & 	$10^{14.3}$ & 	$~30$ & 	$125$ & 	$~360$ & 	$2165$ & 	$~360$ & 	$~549$ & 	$~359$ & 	$~170$  	\\[2pt] 
$1.00$ & 	$1.00$ & 	$10^{14.1}$ & 	$~77$ & 	$~95$ & 	$~313$ & 	$~838$ & 	$~313$ & 	$~~95$ & 	$~263$ & 	$~~~1$  	\\[2pt] 
\hline
\end{tabular}
\label{tbl:mass-cuts}
\end{table*}

The BAHAMAS and MACSIS simulations are both consistent with the \textit{Planck} cosmology, however they use slightly different cosmological parameters, as shown in Table~\ref{tbl:cosmological-parameters}. These differences are not important for this study. 

As shown in~\cite{Barnes2016}, the MACSIS simulations reproduce the mass dependence of the observed gas mass, luminosity and integrated Sunyaev-Zel'dovich signal at $z=0$. They also reproduce the median hot gas profiles of massive galaxy clusters at $z=0$ and $z=1$. 

\subsection{Halo definition and selection}
Haloes are initially identified using the FoF algorithm with linking length $b
=0.2$ times the mean interparticle separation~\citep{Davis1985}. Spherical overdensity masses and radii are determined using the SUBFIND algorithm~\citep{Springel2001}, centred on particles with the minimum gravitational potential in the FoF haloes. %Reference for Mvir is Bryan1998

Only clusters with $M_{200}{\geq}5{\times}10^{13}h^{-1}\mathrm{M}_\odot$ are included in the sample. At redshift zero, the BAHAMAS-HYDRO simulation has 3,298 well resolved galaxy clusters above this mass cut and all 390 MACSIS clusters (in both the HYDRO and DMO simulations) are above this mass cut. Due to its limited box size, the BAHAMAS simulation has very few high-mass clusters; only 9 clusters have masses $M_{200}{\geq}10^{15}h^{-1}\mathrm{M}_\odot$. To ensure the cluster sample is representative, further mass cuts were made to both the BAHAMAS and MACSIS samples. At redshift zero, MACSIS clusters with $M_{200}{\leq} 10^{14.7}h^{-1}\mathrm{M}_\odot$ were found to be underconcentrated, with a median spin parameter of 0.034 in the DMO simulations. Conversely the small fraction of BAHAMAS clusters above this mass cut were found to be overconcentrated, with a median spin parameter of 0.038. For MACSIS, this is a consequence of selecting clusters for resimulation by $M_\mathrm{FoF}$ rather than $M_{200}$. For BAHAMAS, this is likely a statistical fluctuation due to the small number of BAHAMAS clusters above this mass cut. These unrepresentative haloes are removed from the sample. Since only a small number of haloes are removed in this mass cut, it does not affect any of the following results. By making a clean cut in both the MACSIS and BAHAMAS simulations, we can easily separate the two sets of simulations when looking at cluster properties versus $M_{200}$. 

Similar mass cuts are made at the other redshifts considered here, with the mass cuts given in Table~\ref{tbl:mass-cuts}. These mass cuts are used throughout. This table also highlights that the snapshots of the MACSIS and BAHAMAS simulations do not line up perfectly at $z{\neq}0,1$. As a consequence, we only use $z{\approx}0.25,0.5$ when considering redshift independent properties.

\subsection{Relaxation \label{sec:relaxation}}
Since massive galaxy clusters are structures that have collapsed recently, they are dynamic structures which may appear to evolve rapidly. Characterising such systems is difficult and so we define a relaxed sample of clusters, which are expected to be close to dynamical equilibrium and are less affected by recent merger activity.

\begin{figure}
\centering
\includegraphics[width=\linewidth]{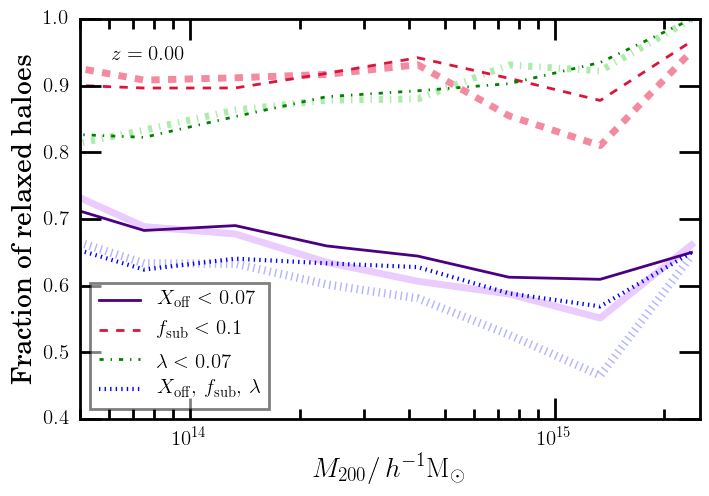}
\caption{The fraction of haloes that are classified as relaxed at $z=0$ using different relaxation criteria: centre of mass offset (solid line), substructure fraction (dashed), the spin parameter (dot dashed) and using all three criteria (dotted). The darker colours show the results for the HYDRO simulations and the lighter coloured, thicker lines are for the DMO simulations.}
\label{fig:fraction-relaxed-haloes}
\end{figure}

Various criteria have been used in the literature to define relaxed haloes, including the centre of mass offset, $X_\mathrm{off}$, the fraction of mass in bound substructures, $f_\mathrm{sub}$, the dimensionless spin parameter, $\lambda$, and the virial ratio~\citep[e.g.][]{Neto2007,Duffy2008,Dutton2014,Klypin2016,Meneghetti2014}. When used in conjunction with other criteria the virial ratio only removes a small number of haloes~\citep{Neto2007}, so we do not use it here. The other parameters are calculated as follows:
\begin{itemize}
\item The centre of mass offset, $X_\mathrm{off}$, is the distance between the minimum of the gravitational potential and the centre of mass of a cluster, divided by the virial radius\footnote{The virial radius, $R_\mathrm{vir}$, is the spherical overdensity mass using $\Delta=\Delta_\mathrm{vir}$, where $\Delta_\mathrm{vir}$ is calculated using the approximation given in~\cite{Bryan1998}}. The centre of mass is calculated using all particles within the virial radius. Haloes with $X_\mathrm{off}<0.07$ are classified as relaxed.
\item The substructure fraction, $f_\mathrm{sub}$, is the fraction of mass within the virial radius that is bound in substructures. Substructures are only included if they contain more than 100 particles and if their centre is not separated from the cluster centre by more than the virial radius. Haloes with $f_\mathrm{sub}<0.1$ are classified as relaxed.
\item The spin parameter, $\lambda$, is calculated for all particles within $R_{200}$. We use the alternative expression for the spin parameter from~\cite{Bullock2001b}. Haloes with $\lambda{<}0.07$ are classified as relaxed.
\end{itemize}
The fractions of haloes classified as relaxed according to these criteria are given as a function of mass in Fig.~\ref{fig:fraction-relaxed-haloes}, with the darker, thinner (lighter, thicker) lines indicating the results for the DMO (HYDRO) simulations. All three criteria show some mass dependence, with the centre of mass offset and the substructure fraction giving fewer relaxed haloes at high masses. At higher masses there should be fewer relaxed haloes, since these structures have only formed recently and are likely the result of recent mergers. The spin parameter criterion does not reflect this, since the fraction of haloes classified as relaxed by this criterion increases as a function of mass. This is likely a consequence of the weak mass dependence of the spin parameter~\citep{Bett2007}, which is discussed in Section~\ref{sec:spins}.

The centre of mass offset criterion removes the largest number of haloes, which is consistent with~\cite{Neto2007} and~\cite{Klypin2016}. Since it is the most stringent criterion, we define relaxed haloes as those where the centre of mass offset, $X_\mathrm{off}<0.07$, unless stated otherwise.

\section{Characterising Massive Galaxy Clusters}

\begin{figure}
\includegraphics[width=0.98\linewidth]{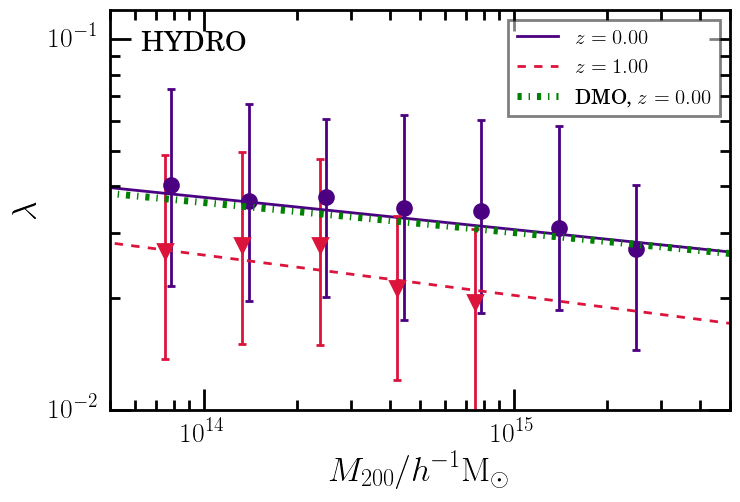}
\includegraphics[width=0.98\linewidth]{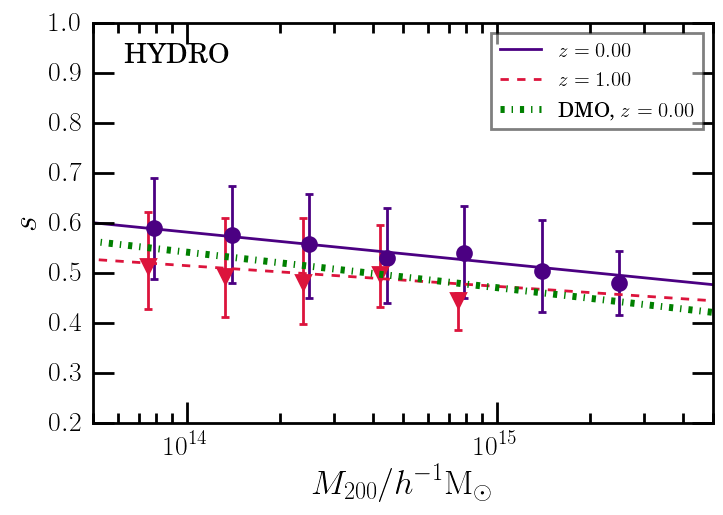}
\includegraphics[width=0.98\linewidth]{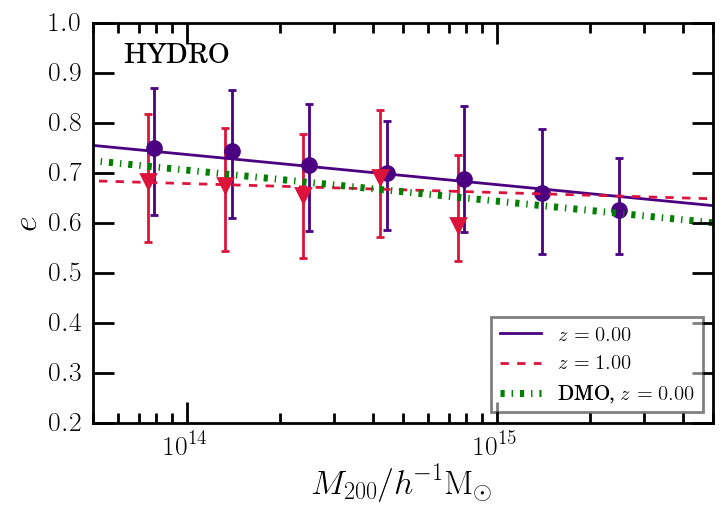}
\caption{The mass dependence of spin, sphericity and elongation in the BAHAMAS-HYDRO and MACSIS-HYDRO simulations at two different redshifts. Markers show the median concentrations in mass bins, with error bars indicating the 16th and 84 percentiles. The lines are fits that are obtained by bootstrap resampling a least squares fit of Equation~\eqref{eq:spin-mass} to individual clusters. The green dot-dashed lines show the best fit relations for the DMO simulations at $z=0$. All three parameters decrease with increasing mass, however the $s-M_{200}$ and $e-M_{200}$ relations get flatter with increasing redshift, whereas the $\lambda-M_{200}$ relation steepens with increasing redshift.}
\label{fig:spin-shape-mass}
\end{figure}

We use three measures to characterise galaxy clusters: the spin, shape and density profile. The latter is quantified using the concentration parameter. 
\begin{table*}
\caption{The mean and standard deviation of the halo spin, $\lambda$, shape parameters $s$ and $e$, and concentration $c_{200}$ at $z=0$ for haloes in the MACSIS and BAHAMAS simulation with $M_{200}{\geq} 5{\times}10^{13}\,h^{-1}\mathrm{M}_\odot$. Errors represent $1\sigma$ confidence intervals, which are determined by bootstrap resampling the sample 1000 times.}
\begin{tabular}{l|c|c|c|c|c|c|c|c|c|c|c|c|}
\hline 
  & \multicolumn{2}{c}{$\log_{10}\lambda$} &\multicolumn{2}{c}{$s=c/a$}&\multicolumn{2}{c}{$e=b/a$}&\multicolumn{2}{c}{$\log_{10}c_{200}$} \\
Sample & Mean &  $\sigma$ & Mean & $\sigma$ &  Mean  & $\sigma$ &  Mean  & $\sigma$  \\
\hline
DMO&$	-1.446^{+0.006}_{-0.006}$&$	0.281^{+0.005}_{-0.005}	$&$0.537^{+0.002}_{-0.002}$&$	0.107^{+0.002}_{-0.002}$&$	0.701^{+0.003}_{-0.003}	$&$0.128^{+0.002}_{-0.002}	$&$0.642^{+0.003}_{-0.004}	$&$0.154^{+0.003}_{-0.003}$ \\[2pt]
DMO (relaxed)	& $ -1.515^{+0.007}_{-0.007}$&$		0.253^{+0.006}_{-0.005}$&$		0.565^{+0.003}_{-0.003}	$&$	0.097^{+0.002}_{-0.002}$&$		0.731^{+0.003}_{-0.003}$&$		0.116^{+0.002}_{-0.002}	$&$	0.700^{+0.003}_{-0.003}$&$	0.119^{+0.003}_{-0.003}	 $ \\[2pt]
HYDRO	&$ -1.434^{+0.007}_{-0.006}$&$	0.278^{+0.005}_{-0.005}$&$	0.576^{+0.002}_{-0.002}	$&$0.105^{+0.002}_{-0.002}	$&$0.732^{+0.003}_{-0.003}	$&$0.123^{+0.002}_{-0.002}	$&$0.601^{+0.003}_{-0.003}	$&$ 0.145^{+0.003}_{-0.003}$ \\[2pt]
HYDRO (relaxed) &$	-1.504^{+0.007}_{-0.007}	$&$ 0.251^{+0.006}_{-0.006}	$&$0.606^{+0.003}_{-0.003}$&$	0.093^{+0.002}_{-0.002}$&$	0.761^{+0.003}_{-0.003}$&$	0.109^{+0.002}_{-0.002}$&$	0.657^{+0.003}_{-0.003}	$&$ 0.109^{+0.004}_{-0.003}$ \\[2pt]
HYDRO, DM &$	-1.410^{+0.007}_{-0.007}	$&$0.280^{+0.005}_{-0.005}$&$	0.546^{+0.003}_{-0.003}	$&$0.109^{+0.002}_{-0.002}	$&$0.714^{+0.003}_{-0.003}$&$	0.129^{+0.002}_{-0.002}$&$	0.621^{+0.003}_{-0.004}$&$	0.151^{+0.003}_{-0.003} $ \\[2pt]
HYDRO, DM (relaxed)	&$ -1.481^{+0.007}_{-0.007}$&$	0.251^{+0.006}_{-0.006}$&$	0.575^{+0.003}_{-0.003}$&$	0.097^{+0.002}_{-0.002}$&$	0.743^{+0.003}_{-0.003}$&$	0.116^{+0.002}_{-0.002}$&$	0.678^{+0.003}_{-0.003}$&$	0.116^{+0.004}_{-0.003} $ \\[2pt]

\hline
\end{tabular}

\label{tbl:param-dist}
\end{table*}

\begin{table*}
\caption{Best fit slope and intercept parameters for the mass dependence of halo spin, $\lambda$, shape parameters, $s$ and $e$, and concentration $c_{200}$, assuming the parameters (as they are listed in the table) are linearly related to $\log_{10}(M_{200}/10^{14}\,h^{-1}\,\mathrm{M}_\odot$). For all but $c_{200}$, the fits are performed for haloes with $M_{200}{\geq} 5{\times}10^{13}\,h^{-1}\mathrm{M}_\odot$. For $c_{200}$, only haloes with $M_{200}{\geq}10^{14}\,h^{-1}\mathrm{M}_\odot$ are used to ensure the density profiles are converged over the radial range $0.05{\leq}r/R_\mathrm{vir}{\leq}1$. Errors represent $1\sigma$ confidence intervals, which are determined by bootstrap resampling the sample 1000 times.}
\begin{tabular}{l|c|c|c|c|c|c|c|c|}%{p{1.65cm}p{0.6cm}p{.6cm}p{.6cm}p{.6cm}p{.6cm}p{.6cm}}
\hline
& \multicolumn{2}{c}{$\log_{10}\lambda$} &\multicolumn{2}{c}{$s=c/a$}&\multicolumn{2}{c}{$e=b/a$}&\multicolumn{2}{c}{$\log_{10}c_{200}$} \\
& Intercept & Slope & Intercept & Slope & Intercept &	Slope & Intercept &	Slope  \\
\hline
$z=0$ &  &  &  &  &  &  &  & \\
DMO & $0.0362^{+0.0004}_{-0.0004}$ & $-0.0810^{+0.0120}_{-0.0120}$ & $0.541^{+0.002}_{-0.002}$ & $-0.071^{+0.004}_{-0.004}$ & $0.705^{+0.002}_{-0.002}$ & $-0.062^{+0.005}_{-0.005}$ & $4.511^{+0.057}_{-0.055}$ & $-0.138^{+0.010}_{-0.009}$\\[2pt]
DMO (relaxed) & $0.0307^{+0.0003}_{-0.0003}$ & $-0.0432^{+0.0134}_{-0.0136}$ & $0.568^{+0.002}_{-0.002}$ & $-0.076^{+0.005}_{-0.005}$ & $0.733^{+0.002}_{-0.002}$ & $-0.071^{+0.006}_{-0.007}$ & $5.195^{+0.058}_{-0.057}$ & $-0.149^{+0.010}_{-0.010}$\\[2pt]
HYDRO & $0.0374^{+0.0004}_{-0.0004}$ & $-0.0873^{+0.0117}_{-0.0113}$ & $0.581^{+0.002}_{-0.002}$ & $-0.062^{+0.004}_{-0.004}$ & $0.736^{+0.002}_{-0.002}$ & $-0.060^{+0.005}_{-0.005}$ & $4.068^{+0.048}_{-0.047}$ & $-0.073^{+0.009}_{-0.009}$\\[2pt]
HYDRO (relaxed) & $0.0316^{+0.0004}_{-0.0004}$ & $-0.0695^{+0.0131}_{-0.0130}$ & $0.610^{+0.002}_{-0.002}$ & $-0.067^{+0.004}_{-0.004}$ & $0.765^{+0.002}_{-0.002}$ & $-0.068^{+0.006}_{-0.006}$ & $4.626^{+0.104}_{-0.108}$ & $-0.074^{+0.008}_{-0.008}$\\[2pt]
 &  &  &  &  &  &  &  & \\[2pt]
$z=1$ &  &  &  &  &  &  &  & \\[2pt]
DMO & $0.0260^{+0.0005}_{-0.0005}$ & $-0.0772^{+0.0269}_{-0.0274}$ & $0.472^{+0.003}_{-0.003}$ & $-0.046^{+0.010}_{-0.010}$ & $0.651^{+0.004}_{-0.004}$ & $-0.016^{+0.013}_{-0.013}$ & $-$ & $-$\\[2pt]
DMO (relaxed) & $0.0206^{+0.0005}_{-0.0005}$ & $-0.0542^{+0.0343}_{-0.0369}$ & $0.496^{+0.004}_{-0.004}$ & $-0.055^{+0.014}_{-0.014}$ & $0.671^{+0.005}_{-0.005}$ & $-0.021^{+0.017}_{-0.018}$ & $-$ & $-$\\[2pt]
HYDRO & $0.0261^{+0.0005}_{-0.0005}$ & $-0.1085^{+0.0276}_{-0.0277}$ & $0.513^{+0.003}_{-0.003}$ & $-0.042^{+0.010}_{-0.010}$ & $0.678^{+0.004}_{-0.004}$ & $-0.019^{+0.013}_{-0.013}$ & $-$ & $-$\\[2pt]
HYDRO (relaxed) & $0.0210^{+0.0005}_{-0.0005}$ & $-0.0680^{+0.0339}_{-0.0343}$ & $0.541^{+0.004}_{-0.004}$ & $-0.054^{+0.013}_{-0.013}$ & $0.700^{+0.005}_{-0.005}$ & $-0.028^{+0.016}_{-0.017}$ & $-$ & $-$\\[2pt]

\hline 
\end{tabular}

\label{tbl:param-mass-fits}
\end{table*}

\subsection{Spins \label{sec:spins}}
The spin parameter, $\lambda$, measures the proportion of energy that is due to the rotation of a cluster. Calculating this parameter requires measuring the total energy of a cluster, which is difficult to define and computationally expensive to compute. Instead, the alternative expression from~\cite{Bullock2001b} is used to gain an estimate of the spin parameter,
\begin{equation}
\lambda = \frac{J}{\sqrt{2}MV_c R}, \label{eq:spin-definition}
\end{equation}
where $J$ is the total angular momentum of matter enclosed within a sphere of radius $R$ and mass $M$, and $V_c$ is the circular velocity at this radius, $V_c=\sqrt{GM/R}$. $\lambda$ is evaluated at $R=R_{200}$ throughout. For the dark matter component in the HYDRO simulations, the spin parameter is calculated using Equation~\ref{eq:spin-definition} with the total mass of dark matter particles within $R_{200}$.

The distribution of spins in the DMO simulations is well fitted by a lognormal distribution, in agreement with~\cite{Bailin2005},~\cite{Bryan2013} and~\cite{Baldi2016} for lower mass haloes. In contrast to ~\cite{Bett2007}, we find no evidence for a longer tail to small values of $\lambda$, however this may be a consequence of the difference in sample size;~\cite{Bett2007} considered ${>}10^6$ haloes.
As Table~\ref{tbl:param-dist} shows, the mean spin parameters are consistent between the DMO and HYDRO simulations. However, the dark matter exhibits a larger mean spin parameter in the HYDRO simulations as compared to the DMO simulations. This is due to a transfer of angular momentum from baryons to the dark matter~\citep{Bett2010,Bryan2013}, which becomes evident by considering the specific angular momentum, $j=J/M$, where $J$ and $M$ are defined in Equation~\ref{eq:spin-definition}. The mean specific angular momentum of the dark matter component increases from $\log_{10}(j/\,h^{-1}\mathrm{Mpc}^2\,\mathrm{s}^{-1})=1.50$ in the DMO simulations to $\log_{10}(j/\,h^{-1}\mathrm{Mpc}^2\,\mathrm{s}^{-1})=1.52$ in the HYDRO simulations, which causes an increase in the spin parameter of the dark matter in the HYDRO simulations.

At $z=0$, selecting only relaxed haloes reduces the mean spin parameter by 15\% in both the DMO and HYDRO simulations, which is consistent with~\cite{Maccio2007},~\cite{Jeeson-Daniel2011} and~\cite{Bryan2013}.

The mass dependence of spins for the HYDRO simulations is shown in the top panel of Fig.~\ref{fig:spin-shape-mass}, where the markers indicate mean values in mass bins. The lines indicate fits to all individual clusters assuming a relation of the form 
\begin{equation}
\log_{10}\lambda= \log_{10}A + B\log_{10}\left(M_{200}/10^{14}h^{-1}\mathrm{M_\odot}\right),  \label{eq:spin-mass}
\end{equation}
where $A$ and $B$ are the best fit parameters. Uncertainties on these parameters are obtained by bootstrap resampling the sample 1000 times.

In agreement with the DMO results from~\cite{Bett2007} and~\cite{Muonoz-Cuartas2011}, spin decreases weakly with increasing mass at all redshifts. The slopes at different redshifts are consistent within the scatter, yet the normalisation decreases with increasing redshift. This is contrary to~\cite{Muonoz-Cuartas2011}, who found a variable slope for lower mass haloes. This may be a consequence of the difference in mass range considered in the studies. We focus on the high mass (${>}5{\times}10^{13}h^{-1}\mathrm{M}_\odot$) end of the relation, whereas~\cite{Muonoz-Cuartas2011} have only a small number of high mass clusters.

As Table~\ref{tbl:param-mass-fits} indicates, the normalisation of the spin-mass relation is slightly larger in the HYDRO simulations than in the DMO simulations. Considering only relaxed haloes (as determined by $X_\mathrm{off}$), reduces the normalisation of the $\lambda{-}M_{200}$ relation at all redshifts by around $15{-}20\%$. The slope of the relation is consistent between the HYDRO and DMO simulations for the full sample. Once only relaxed haloes are selected, we find that the slope is shallower in the DMO simulations. 

\subsection{Shapes}
The shape of a cluster can be characterised by the mass distribution tensor, $\bm{\mathcal{M}}$, or equivalently the inertia tensor, $\bm{I}$~\citep[e.g.][]{Bett2007}. In either of these approaches, the cluster is modelled as a uniform ellipsoid with semi-principal axis lengths $a\geq b\geq c$. The mass distribution tensor of a cluster consisting of $N$ particles is a square matrix with components
\begin{equation}
\mathcal{M}_{ij} = \sum_{k=1}^{N_{200}} m_k r_{k,i}r_{k,j},
\end{equation}
where $m_k$ is the mass of the $k$th particle, $r_{k,i}$ is the $i$th component of the position vector, $\vec{r}_\mathrm{k}$, of the $k$th particle from the centre of the cluster and the sum is over all particles within $R_{200}$. The square roots of the eigenvalues of the matrix $\bm{\mathcal{M}}$ are the lengths of the semi-principal axes of the cluster, $a,b$ and $c$.

The shape of the cluster is parametrised in terms of its sphericity, $s=c/a$, and elongation, $e=b/a$. An idealised spherical structure would have $s=e=1$. Following ~\cite{Bailin2005}, we rescale the axis ratios $s{\to}s^{\sqrt{3}}$ and $e{\to}e^{\sqrt{3}}$ to account for calculating the mass tensor within a spherical region. As discussed in~\cite{Zemp2011},~\cite{Bett2012} and~\cite{Bryan2013}, this simple approach is more comparable with observations than other iterative approaches which measure shape within ellipsoidal regions. 

\begin{figure}
\includegraphics[width=0.95\linewidth]{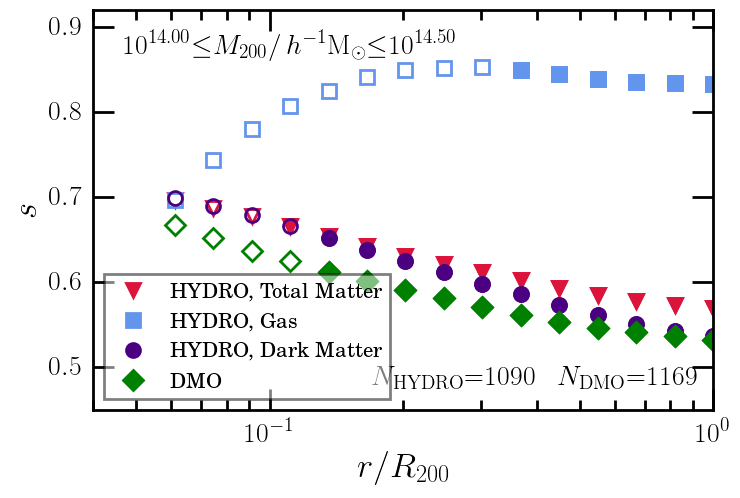} \\
\includegraphics[width=.95\linewidth]{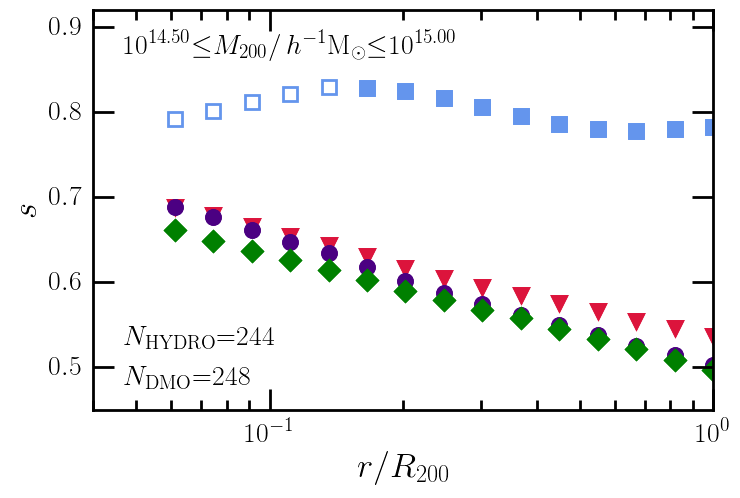}\\
\includegraphics[width=.95\linewidth]{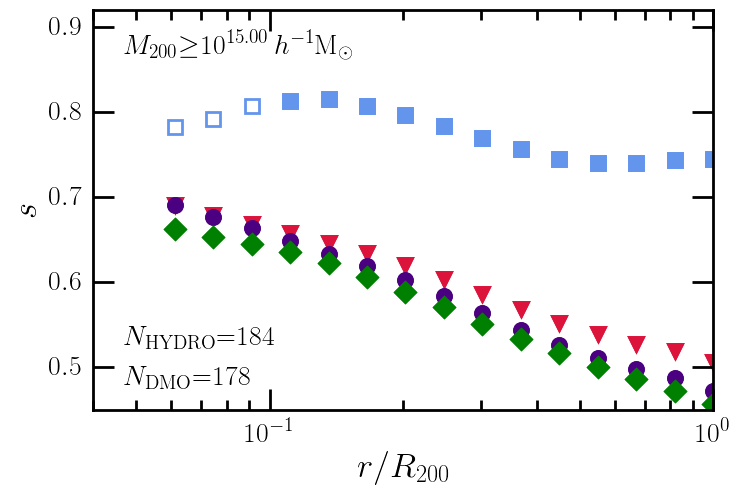}
\caption{The sphericity as a function of radius in mass bins for clusters in the BAHAMAS and MACSIS simulations at $z=0$. The red triangles, light blue squares and purple circles indicate the shapes of the total matter, gas and dark matter distributions in the HYDRO simulations. The green diamonds are for the DMO simulations. Filled markers indicate radii at which the enclosed number of particles is greater than 1000 in each cluster in the bin. Similar trends are apparent in all mass bins; At $r{>}0.2R_{200}$, the shapes of total matter and dark matter distributions in the HYDRO simulations start to diverge as gas starts to contribute significantly. The sphericity profile in the outer regions of the DMO simulations traces the dark matter in the HYDRO simulations, rather than the total matter distribution.}
\label{fig:shape-profiles}
\end{figure}

The distribution of the sphericity in the DMO simulations at $z=0$ is well described by a normal distribution with the mean $\langle s\rangle=0.537$ and standard deviation $\sigma=0.107$, as given in Table~\ref{tbl:param-dist}. In the HYDRO simulations, the mean sphericity increases to $\langle s\rangle=0.576$, whilst the standard deviation does not change significantly. This increase is predominantly due the increased sphericity of gas in the HYDRO simulations, however, dark matter in the HYDRO simulations is also marginally more spherical than in the DMO simulations, with a mean sphericity $\langle s\rangle=0.546$ in the HYDRO simulations. This difference is also present in the elongation.

Sphericity and elongation as a function of mass for all particles in MACSIS and BAHAMAS clusters in the HYDRO simulations are shown in the middle and bottom panels of Fig.~\ref{fig:spin-shape-mass}. Again, markers indicate median values in mass bins, with error bars showing the $1\sigma$ percentiles. The lines indicate best fit relations of the form given in Equation~\ref{eq:spin-mass}, replacing $\log_{10}\lambda$ for $s$ or $e$. The general trend of sphericity and elongation decreasing with increasing cluster mass is in agreement with~\cite{Maccio2008, Muonoz-Cuartas2011} and~\cite{Bryan2013}, indicating that more massive clusters form more extended, aspherical structures. ~\cite{Bryan2013} consider a number of hydrodynamic models, and find the model most relevant to this work (their AGN simulation) exhibits a steeper mass dependence in the $s-M_{200}$ relation with a slope of $-0.078$ at $z=0$. However in the same work it is demonstrated that the relation between halo shape and mass is model dependent, with the slope varying from $-0.034$ to $-0.078$ at $z=0$ for different models. 

The bottom panel of Fig.~\ref{fig:spin-shape-mass} shows the elongation ($e=b/a$) as a function of mass for the HYDRO simulations. The mass dependence of the elongation is weaker than for the sphericity, which suggests that as clusters acquire mass, they preferentially collapse along their shortest axis.

As is evident from Table~\ref{tbl:param-mass-fits}, the normalisation of the $s{-}M_{200}$ relation is around 7\% higher in the HYDRO simulations compared to the DMO simulations at $z=0$. Similarly, the normalisation of the $e{-}M_{200}$ relation is ${\approx}4\%$ higher in the HYDRO simulations. We find the slope of the $s{-}M_{200}$ relation to be steeper by ${\approx}15\%$ in the DMO simulations at $z=0$, although the errors on the slopes are ${\approx}5{-}6\%$, suggesting that a wider mass range is needed to constrain this difference fully. The slopes of the $e{-}M_{200}$ relations in the HYDRO and DMO simulations are consistent with each other.

Table~\ref{tbl:param-mass-fits} also gives the $s{-}M_{200}$ relation for clusters classified as relaxed by $X_\mathrm{off}$. Relaxed clusters are more spherical, with a 5\% increase in the intercept of the $s-M_{200}$ relation at $z=0$. The slope of the $s-M_{200}$ relations for relaxed clusters are consistent with that for the full sample. The trends in the $e-M_{200}$ relation mirror this, with a 4\% increase in the normalisation of the $e-M_{200}$ relation for relaxed haloes and no significant effect on the slope.

Fig.~\ref{fig:shape-profiles} shows the variation of the sphericity with radius for clusters in mass bins in both the HYDRO and DMO simulations. At each radius the sphericity is calculated using all particles enclosed within a sphere of that radius. In both the HYDRO and DMO simulations, the sphericity of the total matter distribution decreases as a function of radius, in agreement with existing work~\citep{Hopkins2005}. Notably, the sphericity profile for the dark matter in the DMO simulations traces the dark matter in the HYDRO simulations in the outer regions. In the central region (considering only radii containing at least 1000 particles), the DMO profiles get shallower in highest mass bin, whilst the dark matter and total matter profiles in the HYDRO simulations do not. As a consequence, clusters in the HYDRO simulations are more spherical in their central regions than DMO clusters, which is likely to be a consequence of the contraction of dark matter in the cluster centres. In the HYDRO simulations, the dark matter dominates the shape of the total matter distribution in the central region, but the contribution of the gas to the total matter distribution becomes significant at $r{>}0.2R_{200}$, when the sphericity profiles of the total matter distribution and dark matter distributions start to diverge.

At $r{<}0.1R_{200}$ in the lowest mass bin, the sphericity profiles in the DMO and the HYDRO simulations seem to reconverge, however a higher resolution study is needed to confirm this since the clusters in this study have an insufficient number of particles for their shape measurements to be well converged there. For the same reason, we cannot comment on the shape of the stellar mass distribution in this study.

\subsection{Density profiles and concentrations}
\subsubsection{The Impact of Baryons on Cluster Profiles ~\label{sec:3d-profs-baryons}}

\begin{figure}
\centering
\includegraphics[width=\linewidth]{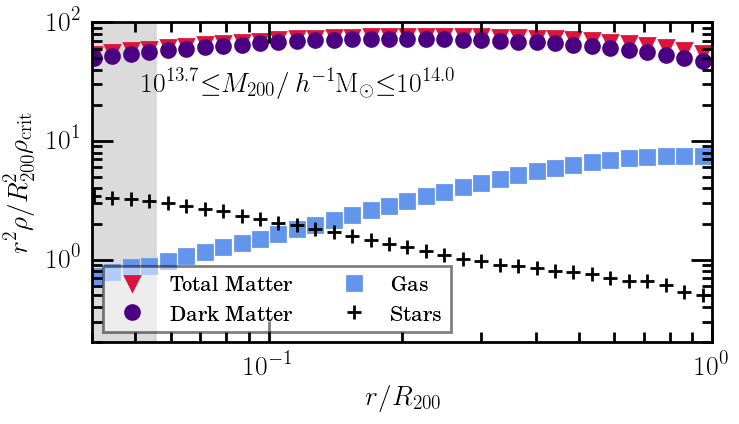}
\includegraphics[width=\linewidth]{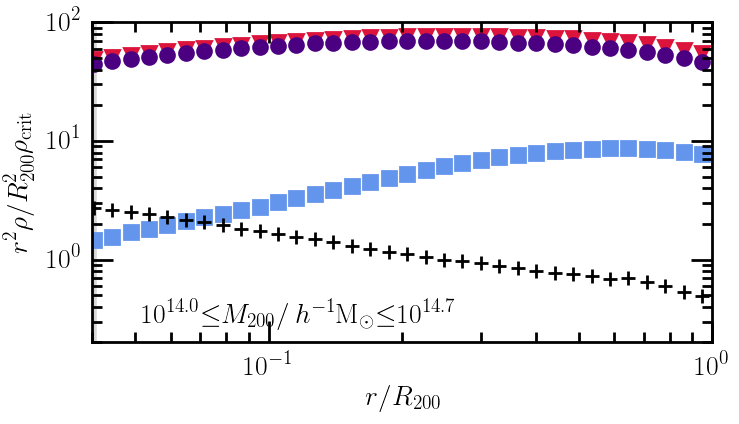}
\includegraphics[width=\linewidth]{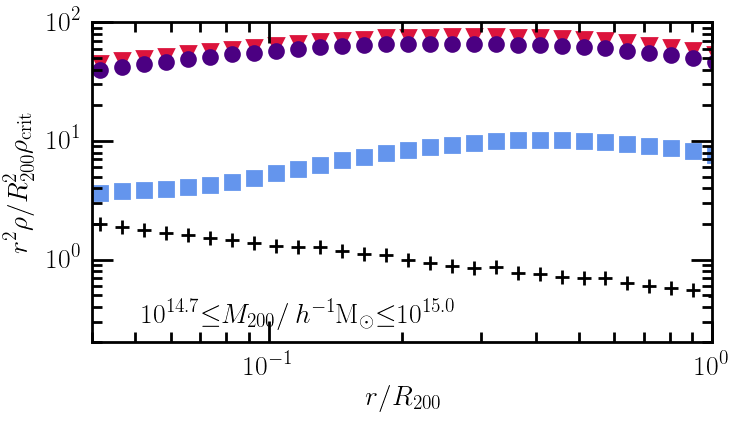}
\includegraphics[width=\linewidth]{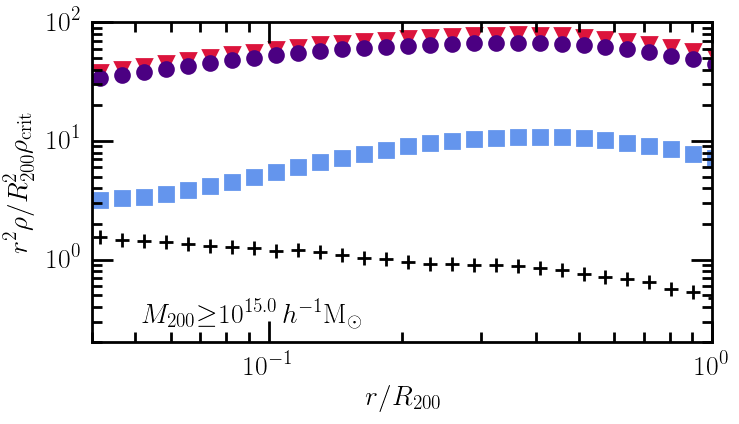}
\caption{The mean density profiles for the gas (blue squares), stars (black crosses), dark matter (purple circles) and total matter (red triangles) for clusters stacked by cluster mass at $z=0$. The top two figures show profiles for clusters in the HYDRO-BAHAMAS simulation and the bottom two panels show profiles of clusters in the HYDRO-MACSIS simulations. The shaded grey region indicates the largest convergence radius in each mass bin. From top to bottom, each bin contains 2058 (0), 1335 (0), 0 (142) and 0 (189) BAHAMAS (MACSIS) clusters respectively.}
\label{fig:meanprofs-red0}
\end{figure}

Density profiles are obtained for clusters by binning particles in 50 equally spaced logarithmic bins between $10^{-2}\mathrm{\leq}r/R_\mathrm{vir}{\leq}1$. Fig.~\ref{fig:meanprofs-red0} shows the mean density profiles for clusters in the HYDRO simulations stacked in mass bins. At all radii considered here, the total matter density profile (red triangles) is dominated by the dark matter component (purples circles). Considering only the baryonic component, stars (black crosses) dominate in the inner region of lower mass clusters, with gas dominating outside of that region. The radius at which stars begin to dominate is neither constant nor a fixed fraction of $R_{200}$. For the most massive clusters ($M_{200}{\geq}10^{15}\,h^{-1}\,\mathrm{M}_\odot$), the gas component dominates over the stellar component at all plotted radii. The shapes of the mean stellar and gas density profiles are consistent with the shapes of mean profiles for haloes with masses greater than $10^{13}h^{-1}\mathrm{M}_\odot$ in~\cite{Schaller2015}.

\begin{figure}
\centering
\includegraphics[width=\linewidth]{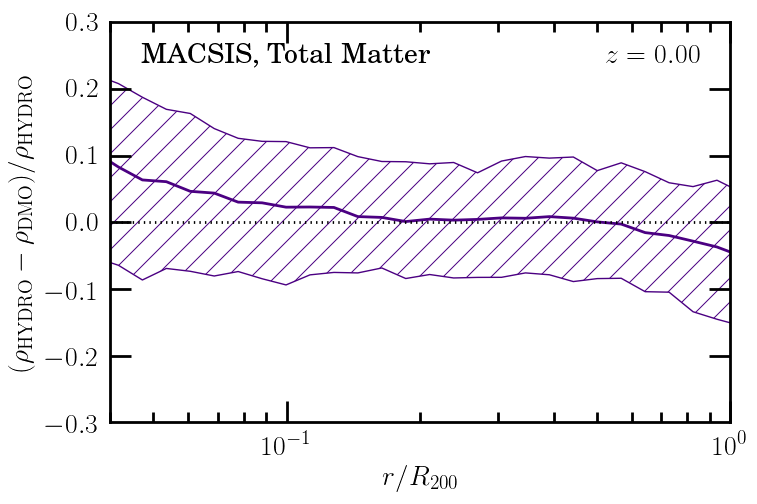}
\includegraphics[width=\linewidth]{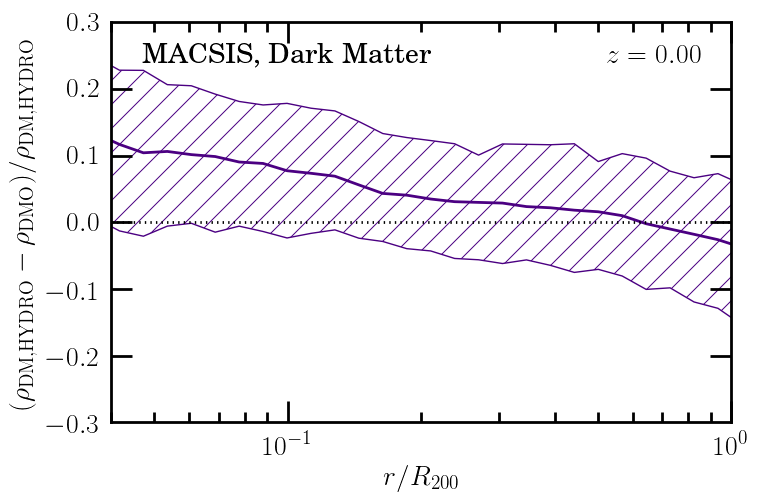}
\caption{In the top panel the solid purple line is the median fractional difference in the total matter density profiles of matched clusters in the DMO and HYDRO MACSIS simulations at $z=0$. These clusters span the mass range $10^{14.7}{\leq}M_{200}/h^{-1}\mathrm{M}_\odot{\leq}10^{15.6}$. The hatched purple region shows the 16th to 84th percentiles.  The bottom panel shows the fractional difference in the dark matter density profiles for matched haloes, where the dark matter density profile in the DMO simulations has been rescaled by the a factor of $\Omega_\mathrm{DM}/\Omega_\mathrm{m}$, where $\Omega_\mathrm{DM}$ is the dark matter fraction and $\Omega_\mathrm{m}$ is the total matter fraction.}
\label{fig:macsis-dmo-hydro-profdiff}
\end{figure}

Since the MACSIS sample consists of 390 individual clusters that have been simulated both as DMO and HYDRO clusters, they are ideal for studying the impact of baryons on the dark matter profile. The top panel of Fig.~\ref{fig:macsis-dmo-hydro-profdiff} shows the mean fractional difference between the total matter density profile in the DMO and HYDRO simulations at $z=0$, where clusters have been individually matched. We see that the density profiles are more concentrated in the HYDRO simulations, with an increase in the density profile at small radii and a decrease at $r{\approx}R_{200}$ compared to the DMO simulations. This difference is not simply due to the baryonic component condensing at the cluster centre; it is also present in the dark matter distribution, as can be seen from the bottom panel of Figure~\ref{fig:macsis-dmo-hydro-profdiff}. Since the clusters considered in this figure all have masses $M_{200}{\geq}10^{14.7}\,h^{-1}\mathrm{M}_\odot$, our results show the impact of baryons on the density profiles of clusters in this mass range. However, this is consistent with previous works looking at less massive structures that have found that the inclusion of baryonic effects leads to a contraction of the inner halo, causing an increase in the dark matter profile at small radii~\citep[e.g.][]{Duffy2010,Schaller2015}.

\subsubsection{NFW or Einasto?}
The density profiles of dark matter haloes are commonly fitted by the two-parameter NFW model, proposed by~\cite{Navarro1997}, 
\begin{equation}
\rho_\mathrm{NFW}(r)=\frac{\rho_\mathrm{crit}\delta_c}{(r/r_{-2})(1+r/r_{-2})^2}
\end{equation}
which is characterised by an overdensity, $\delta_c$ and a scale radius, $r_{-2}$. The scale radius is the radius at which the density profile has an isothermal slope. However, numerous authors have found haloes have a steeper than NFW slope at small radii~\citep{Moore1998,Jing2000a,Fukushige2001}, whilst others have found a shallower slope~\citep{Navarro2004,Merritt2006a}, which suggests that a model with a variable inner slope may be more appropriate.~\cite{Gao2008},~\cite{Dutton2014} and~\cite{Klypin2016} have found that dark matter density profiles more closely follow the Einasto profile~\citep{Einasto1965}:
\begin{equation}
\rho(r) = \delta_c \rho_\mathrm{crit}\exp\left\{-\frac{2}{\alpha} \left[ \left( \frac{r}{r_{-2}}\right)^\alpha -1 \right]\right\}, \label{eq:einasto}
\end{equation}
which has a logarithmic slope parametrised by $\alpha$.

For clusters with $M_{200}{\geq}10^{14}\,h^{-1}\mathrm{M}_\odot$, best fit cluster profiles are obtained by fitting profiles in the radial range $0.05 {\leq} r/R_\mathrm{vir} {\leq} 1$. This mass cut is made to ensure that the convergence radius~\citep[calculated following][]{Power2003} is always within the inner fitting radius. The model profile parameters are adjusted to minimise 
\begin{equation}
\rho_\mathrm{rms}= \frac{1}{N_\mathrm{dof}} \sum_{i}^{N_\mathrm{bins}} \left[\log_{10}\rho_i - \log_{10} \rho_\mathrm{model}(\vec{p})\right]^2, \label{eq:rho-leastsquares}
\end{equation}
where $N_\mathrm{dof}$ is the number of degrees of freedom (e.g. $N_\mathrm{dof}=N_\mathrm{bins}-2$ for the NFW profile), $\rho_i$ is the density in radial bin $i$ and $\vec{p}$ is the vector of parameters: $\vec{p}=(r_{-2},\delta_c)$ for an NFW profile and $\vec{p}=(r_{-2},\delta_c,\alpha)$ for an Einasto profile.

\begin{figure}
\centering
\includegraphics[width=\linewidth]{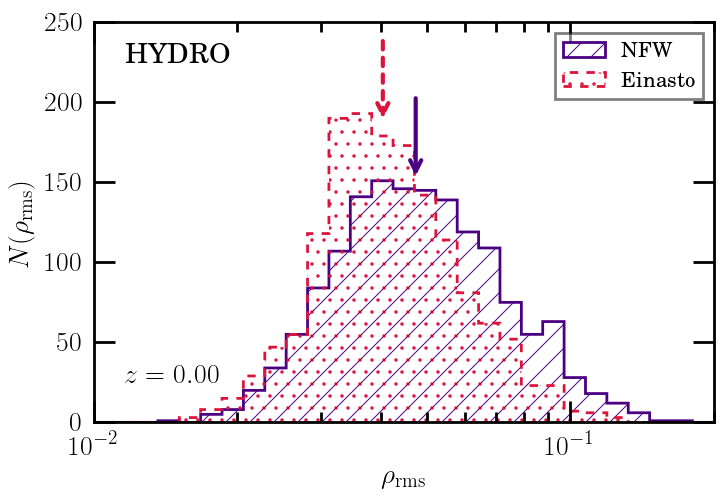}
\includegraphics[width=\linewidth]{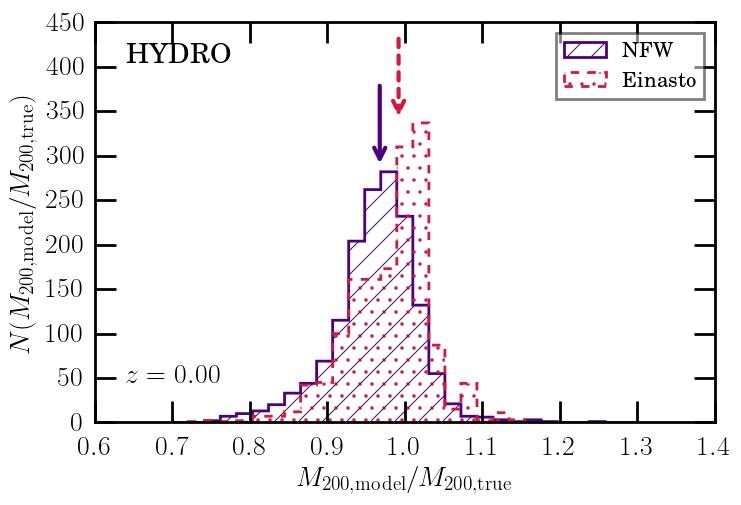}
\caption{The top panel shows the goodness of fit for NFW (purple, diagonal hatching) and Einasto (pink, dotted hatching) fits to clusters in the HYDRO simulations with $M_{200}{\geq}10^{14}\,h^{-1}\mathrm{M}_\odot$ at $z=0$. The arrows indicate median values for the NFW (purple) and Einasto (pink) models respectively. The bottom panel shows the mass inferred from the best fit NFW and Einasto profiles. Einasto profiles provide a better fit to the density profiles of clusters and on average provide better estimates of cluster masses.}
\label{fig:einasto-nfw-3d}
\end{figure}

The top panel in Fig.~\ref{fig:einasto-nfw-3d} shows the goodness of fit (defined in Equation~\ref{eq:rho-leastsquares}) for NFW and Einasto fits to clusters in the HYDRO simulations. For the NFW profile, the goodness of fits are on average larger, with $\mathrm{median}(\rho_\mathrm{rms})=0.047{\pm}0.001$ as compared to $0.040{\pm}0.001$ for the Einasto model, which indicates the NFW model is a slightly poorer fit to cluster profiles than the Einasto model. The goodness of fits for the NFW profile also exhibit a larger scatter. These results are echoed in the DMO simulations (not shown) with $\mathrm{median}(\rho_\mathrm{rms})=0.048{\pm}0.001$ for the NFW model and $0.041{\pm}0.001$ for the Einasto model. 

The bottom panel of Fig.~\ref{fig:einasto-nfw-3d} shows mass estimates obtained from fits to spherically averaged density profiles in the HYDRO simulations. Both the NFW and Einasto models slightly underpredict cluster masses, with $\mathrm{median}(M_{200,\mathrm{model}}/M_\mathrm{200})=0.968^{+0.002}_{-0.001}$ and $0.992^{+0.001}_{-0.001}$ for the NFW and Einasto models respectively. A similar difference is present in the DMO simulations, in which $\mathrm{median}(M_{200,\mathrm{model}}/M_\mathrm{200})=0.976^{+0.002}_{-0.002}$ for the NFW model and $0.992^{+0.001}_{-0.002}$ for the Einasto model. The slight improvement of the Einasto model in reproducing cluster masses over the NFW model is a consequence of the better fit the Einasto model provides to cluster mass profiles.

\subsubsection{The concentration-mass relation}
\begin{figure}
\includegraphics[width=\linewidth]{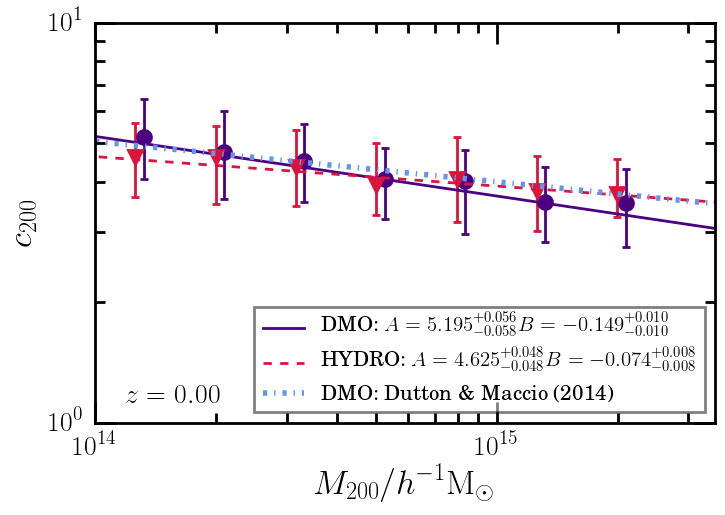} 

\caption{The mass dependence of cluster concentration for the total matter distribution in  relaxed BAHAMAS and MACSIS HYDRO (red triangles) and DMO (purple circles) clusters at two different redshifts. Concentrations are obtained by fitting NFW profiles to the total matter density profiles of clusters over the radial range $0.05{\leq}r/R_\mathrm{vir}{\leq}1.$ Markers show the median concentrations in mass bins, with error bars indicating the 16th to 84 percentiles. The lines are fits that are obtained by bootstrap resampling a least squares fit of Equation~\ref{eq:c200-m200-formula} to individual clusters. The concentration-mass relation from~\protect\cite{Dutton2014} is shown in light blue. DMO clusters have higher concentrations at low masses and exhibit a stronger mass dependence.}
\label{fig:c200-nfw-mass}
\end{figure}

Fig.~\ref{fig:c200-nfw-mass} shows concentration as a function of mass for the total matter distribution in relaxed clusters in the DMO and HYDRO simulations Concentrations are obtained by fitting two-parameter NFW profiles. The relationship between concentration and mass is well fit by a power-law,
\begin{equation}
c_{200}=A\,\left(\frac{M_{200}}{10^{14}h^{-1}\mathrm{M}_\odot}\right)^{B},\label{eq:c200-m200-formula}
\end{equation}
so that in Fig.~\ref{fig:c200-nfw-mass}, $B$ is the slope. The best fit parameters are given in Table~\ref{tbl:param-mass-fits}, with the uncertainties on the fit parameters obtained through bootstrap resampling the fit 1000 times. The best fit concentration-mass relation for the DMO simulation exhibits a steeper slope than that found in literature ~\citep{Neto2007,Duffy2008,Dutton2014}, however Fig.~\ref{fig:c200-nfw-mass} illustrates that the data are consistent with the results of~\cite{Dutton2014}, which uses the \textit{Planck} cosmology. The difference between the concentration-mass relation presented here and that found in wider literature is not surprising since this analysis is limited only to large masses ($M_{200}{\geq}10^{14}h^{-1}\mathrm{M}_\odot$), which have not been extensively studied in other works. 

\vspace{20pt}
In summary, low mass clusters in the HYDRO simulations are more spherical and more centrally concentrated than their DMO counterparts. This is a consequence of both the condensation of baryons in the cluster centre and the contraction of the dark matter halo in the presence of baryons. It is more significant in high mass clusters, which leads to a flatter concentration-mass relation in the HYDRO simulations. The density profiles of clusters in both the HYDRO and DMO simulations are well fit by the NFW profile, however the Einasto model provides a marginally better fit and gives less biased mass estimates.

\section{Cluster weak lensing}
\begin{figure*}
\centering
\includegraphics[width=\linewidth]{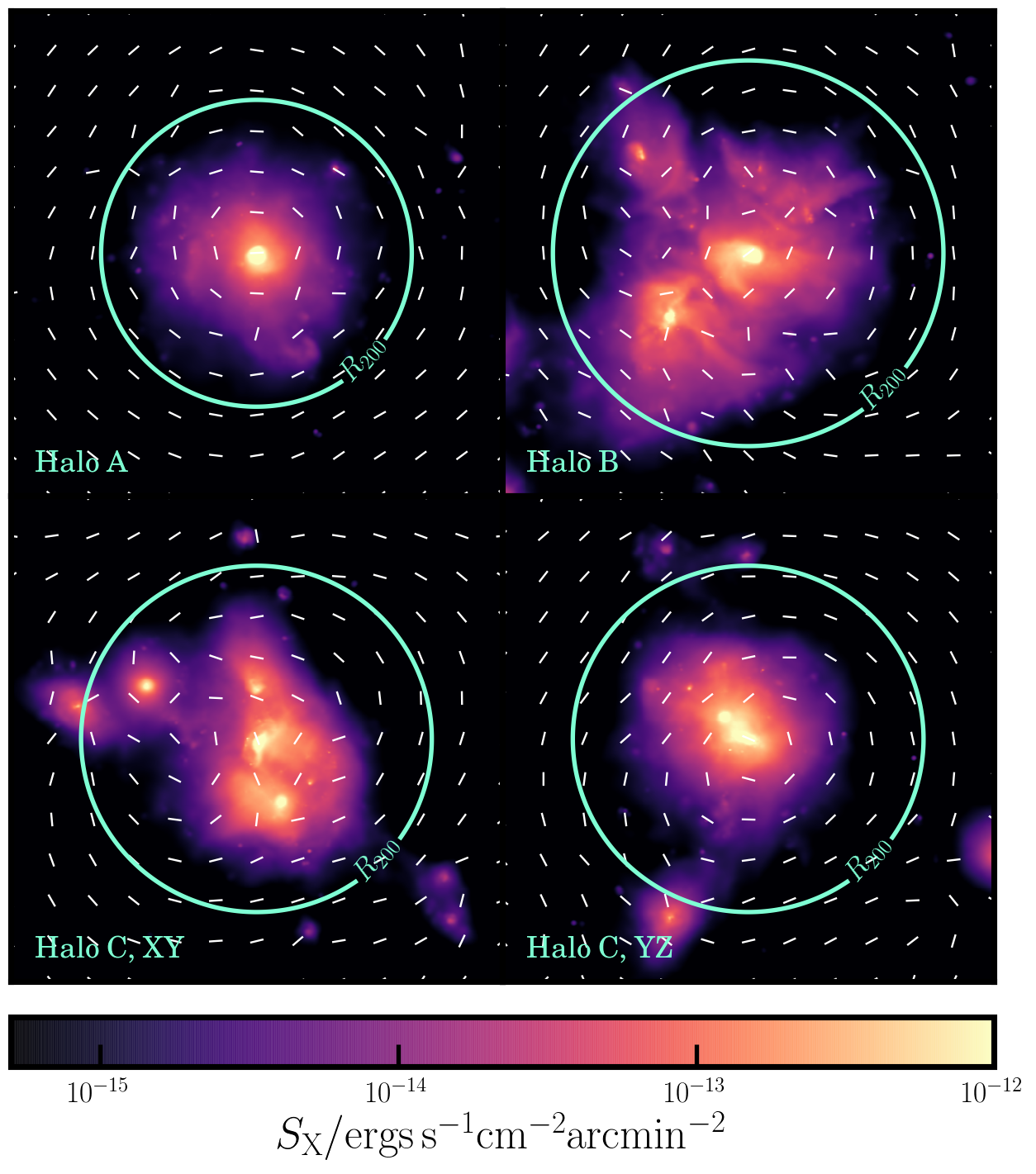}
\caption{The X-ray surface brightness, $S_X$ of four MACSIS clusters at $z=0.24$ with quivers representing the weak lensing shear field. Each image is centred on the minimum gravitational potential for the cluster and is $6\,h^{-1}\mathrm{Mpc}\,{\times}\,6\,h^{-1}\mathrm{Mpc}$ across, with a projection depth of $10\,R_{200}$. The MACSIS sample contains a wide range of relaxed (for example the top left cluster) and unrelaxed clusters (top right). The bottom two images show two orthogonal projections of the same cluster. In the image on the bottom left, the cluster appears to be extended with multiple X-ray peaks, which may lead to it being classed as morphologically unrelaxed in observations. In contrast, the image on the bottom right, whilst still containing multiple X-ray peaks, is more spherically symmetric.}
\label{fig:xray-shear-maps}
\end{figure*}
The use of galaxy clusters as cosmological probes requires well constrained galaxy cluster masses. Cluster weak lensing, which measures the statistical distortion of background galaxies due to the mass of the intervening cluster, is touted as a largely unbiased technique for measuring cluster masses. Furthermore, the shear profiles of galaxy clusters are also used to test $\Lambda$CDM and theories of modified gravity~\citep[e.g.][]{Okabe2013,Wilcox2015}. The shear profiles and weak lensing mass estimates of galaxy clusters have been studied extensively in dark matter only simulations~\citep{Becker2011,Bahe2012}, however the impact of baryons on the projected mass distribution is not so well understood. 

Weak lensing studies measure the shape distortion of background galaxies, which is quantified in the reduced shear,
\begin{equation}
g=\frac{\gamma}{1-\kappa},
\end{equation}
in which $\gamma$ is the shear and $\kappa$ the convergence. The shear describes the tidal gravitational force and has two components, $\gamma=\gamma_1 + i \gamma_2$. The convergence describes the isotropic focussing of light and is proportional to the projected surface density of the lens, $\Sigma$,
\begin{equation}
\kappa = \frac{\Sigma}{\Sigma_\mathrm{crit}}, \label{eq:convergence}
\end{equation}
where $\Sigma(R)$ is the integral of the three dimensional density profile along the line-of-sight, 
\begin{equation}
\Sigma(R) = 2\int\limits_0^\infty \rho(r=\sqrt{R^2+z^2})dz, \label{eq:surface-density}
\end{equation}
and $\Sigma_\mathrm{crit}$ is the critical surface density,
\begin{equation}
\Sigma_\mathrm{crit} \equiv \frac{1}{4\pi G}\frac{D_s}{D_d D_{ds}},
\end{equation}
where $D_s$, $D_d$ and $D_{ds}$ are the angular distances between the observer and source galaxies, observer and lens and lens and source galaxies respectively~\citep[e.g.][]{Wright2000}. We ignore the effect of shape noise, which is noise due to averaging over a finite number of source galaxies within each pixel. 

Weak lensing studies of clusters probe only shape distortions tangential to the line from the projected cluster centre. The tangential component of the shear is 
\begin{equation}
\gamma_t = \mathrm{Re}\left[\gamma \mathrm{e}^{-2i\phi}\right],
\end{equation}
where $\phi$ is the polar angle of the cluster~\citep[e.g.][]{Bartelmann2001}.

\subsection{Weak lensing shear and X-ray surface brightness maps}

\begin{figure}
\centering
\includegraphics[width=\linewidth]{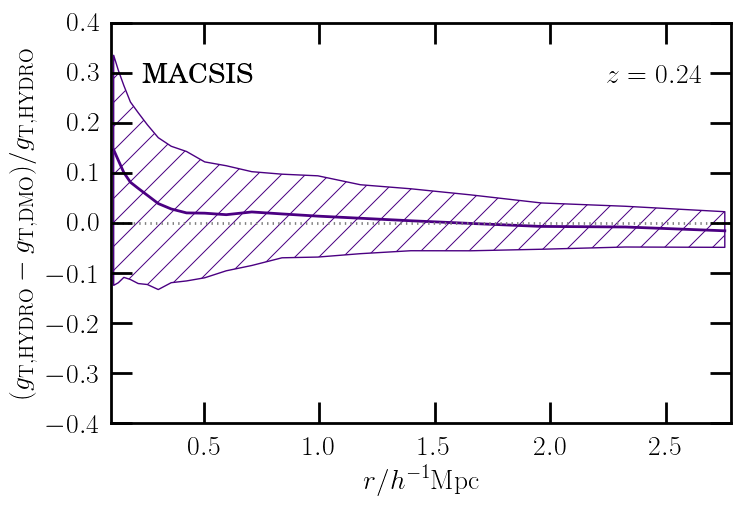}
\includegraphics[width=\linewidth]{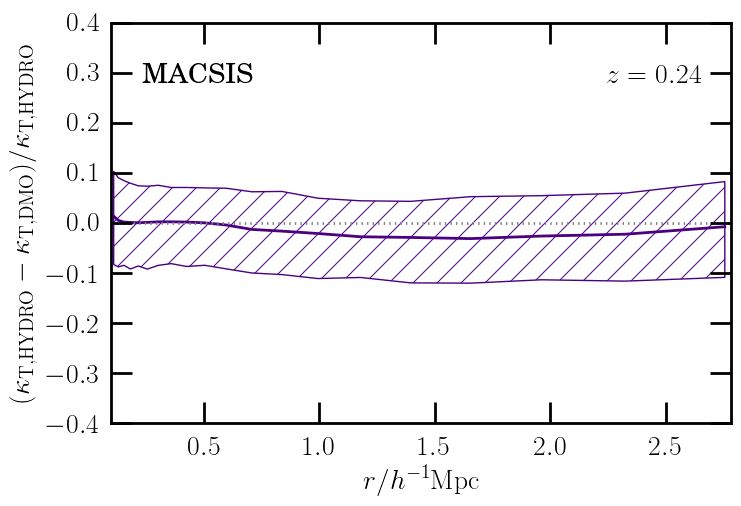}
\caption{In the top panel the solid purple line is the median fractional difference in the reduced tangential shear profiles, $g_\mathrm{T}(r)$ of matched clusters in the DMO and HYDRO MACSIS simulations at $z=0.24$. The hatched purple region shows the 16th and 84th percentiles. The bottom panel shows the fractional difference in the convergence profiles of matched haloes at the same redshift. Baryons have a stronger impact in the central regions of the reduced shear profile than is apparent in the convergence profile.}
\label{fig:wl-shearprof-baryons}
\end{figure}
Surface density maps are produced for each cluster in the BAHAMAS and MACSIS simulations by selecting all particles within a radius of $5R_{200}$ of the cluster centre and projecting these along the desired line of sight. Reducing the selection region to within a radius of $3R_{200}$ of the cluster centre doesn't affect the results presented here, however reducing this radius further changes the results. Particles are then smoothed to a 2D grid with cell width $10\,h^{-1}\mathrm{kpc}$ using smoothed particle hydrodynamics (SPH) smoothing with 48 neighbours. Gas, stars and dark matter are smoothed separately, and the resulting maps are summed to give a total matter mass map. Three orthogonal projections were taken of each cluster, one along each axis of the simulation box.

Convergence maps are obtained by dividing surface density maps by $\Sigma_\mathrm{crit}$. The shear is related to the convergence through their Fourier transforms~\citep[e.g.][]{Schneider2006},
\begin{equation}
\tilde{\gamma}=\left(\frac{\hat{k}_x^2 -\hat{k}_y^2}{\hat{k}_x^2+k_y^2}+i \frac{2\hat{k}_x\hat{k}_y}{\hat{k}_x^2+\hat{k}_y^2}\right)\tilde{\kappa}, \label{eq:shear-FT}
\end{equation}
where $\tilde{\gamma}$ and $\tilde{\kappa}$ are the Fourier transforms of the shear and convergence respectively and $\hat{k}_x$ and $\hat{k}_y$ are wavenumbers. The source redshift is taken to be $z=1$ throughout this work.

We also compute X-ray surface brightness maps to compare the gas and total matter distributions. These are produced as follows. The X-ray luminosity for each particle is obtained using the cooling function calculated using the Astrophysical Plasma Emission Code~\citep[APEC;][]{Smith2001} with updated atomic data and calculations from the AtomDB v2.0.2~\citep{Foster2012}. We use the element abundances that are tracked in the simulation (H, He, C, N, O, Ne, Mg, Si, S, Ca and Fe). The X-ray surface brightness is calculated from the luminosity by dividing through by the angular area of each pixel. The distribution of particle luminosities within $5R_{200}$ is then projected along one axis and smoothed using SPH smoothing to give a 2D map of the X-ray emission. Further details of this approach are given in~\cite{Barnes2016}.

Fig.~\ref{fig:xray-shear-maps} shows the shear field (tick marks) of four MACSIS clusters at $z=0.24$, with X-ray surface brightness in the background. The top left image in Fig.~\ref{fig:xray-shear-maps} is of a dynamically relaxed cluster with mass $M_{200}=1{\times}10^{15}h^{-1}\mathrm{M}_\odot$, with a roughly symmetrical shear field and only one X-ray peak. In contrast, the image on the top right is of a merger with $M_{200}=2{\times}10^{15}h^{-1}\mathrm{M}_\odot$, which shows how the presence of substructure disturbs the shear field. The bottom two images are two orthogonal projections of the same $M_{200}=1.5{\times}10^{15}h^{-1}\mathrm{M}_\odot$ cluster. Considering only the emission within $R_{200}$, the X-ray emission of the YZ (right panel) projection appears relatively relaxed and the shear field is roughly symmetrical. However, in the XY projection (left), we see multiple X-ray peaks and a perturbed shear field, illustrating how one cluster can look drastically different in different projections. 

\subsection{Weak lensing profiles and mass estimates \label{sec:wl-profiles}}

In observations, galaxy cluster masses are obtained from a reduced tangential shear map by first calculating a shear profile and then fitting it with a model profile, from which a mass can be inferred. We obtain reduced tangential shear profiles for each cluster by finding the mean reduced shear in 20 logarithmically spaced bins in the range  $0.1{\leq} r/h^{-1}\mathrm{Mpc}{\leq}3$.
\begin{figure}
\includegraphics[width=\linewidth]{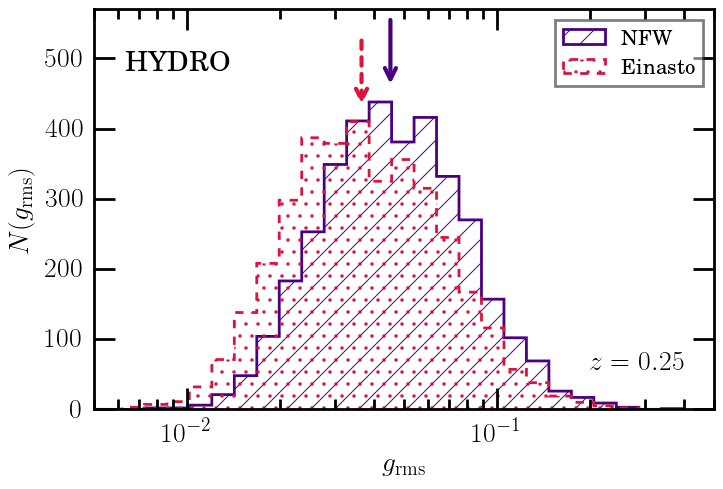}
\includegraphics[width=\linewidth]{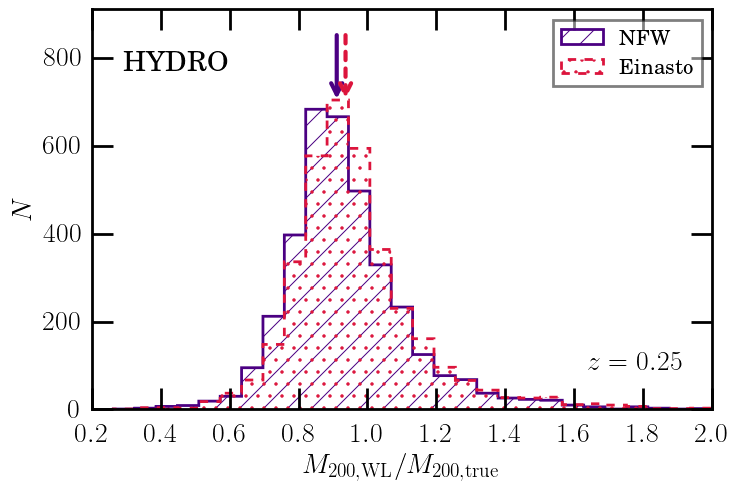}
\caption{The top panel shows the goodness of fit for the best fit NFW (purple, diagonal hatching) and Einasto (pink, dotted hatching) profiles to the reduced tangential shear profiles of clusters in the HYDRO simulations with $M_{200}{\geq}10^{14}\,h^{-1}\mathrm{M}_\odot$ at $z=0.25$. The bottom panel shows masses inferred from these fits in units of the true mass. As in 3D, Einasto profiles provide a better fit to the density profiles of clusters and on average provide slightly better estimates of cluster masses.}
\label{fig:einasto-nfw-2d}
\end{figure}
Both the NFW and Einasto models assume spherical symmetry. For an axisymmetric halo, the radial dependence of the tangential shear is~\citep[e.g.][]{Wright2000}
\begin{equation}
\gamma (r) = \frac{\bar{\Sigma}(<r) - \Sigma(r)}{\Sigma_\mathrm{crit}}, \label{eq:axisymmetric-shear}
\end{equation}
where $\bar{\Sigma}(<r)$ is the mean surface mass density of the halo within a radius $r$, 
\begin{equation}
\bar{\Sigma}(<r) = \frac{2}{r^2}\int\limits_0^r x \Sigma(x) dx. \label{eq:mean-surface-density}
\end{equation}
To obtain the shear of an NFW halo, Equation~\ref{eq:mean-surface-density} is numerically integrated using the analytic form for $\Sigma(r)$ from~\cite{Bahe2012}. This is then substituted into Equation~\ref{eq:axisymmetric-shear}. A similar process is used for Einasto profiles, however in the absence of an analytic form for $\Sigma(r)$ using a truncated line of sight, Equation~\ref{eq:surface-density} is numerically integrated to obtain $\Sigma(r)$ for an Einasto halo. 

The best fit model is found by minimising
\begin{equation}
g_\mathrm{rms} = \frac{1}{N_\mathrm{dof}} \sum_i^{N_\mathrm{bins}} \left[\log_{10} g_{\mathrm{T},i}-\log_{10}g_\mathrm{T,model}(r,\vec{p})\right]^2,
\end{equation}
where $N_\mathrm{dof}$ is the number of degrees of freedom, $g_{\mathrm{T},i}$ is the reduced tangential shear measured in the $i$th shell and $\vec{p}$ is the vector of fit parameters; $\vec{p}=(r_s,\delta_c)$ for an NFW profile and $\vec{p}=(r_s,\delta_c,\alpha)$ for an Einasto profile. Given the best fit parameters for a particular model, an estimate of $R_{200}$ is obtained by solving the equation
\begin{equation}
\int\limits^{R_\mathrm{200,WL}}_{0} \rho(r,\vec{p})r^2dr = \frac{200}{3}\rho_\mathrm{crit}(z)R_\mathrm{200,WL}^2 \label{eq:wl-r200}
\end{equation}
for $R_\mathrm{200,WL}$. The cluster mass estimate, $M_\mathrm{200,WL}$, is then given by 
\begin{equation}
M_\mathrm{200,WL}=\frac{4\pi}{3}200 \rho_\mathrm{crit}(z) R_\mathrm{200,WL}^3. \label{eq:wl-m200}
\end{equation}

\begin{figure}
\centering
\includegraphics[width=\linewidth]{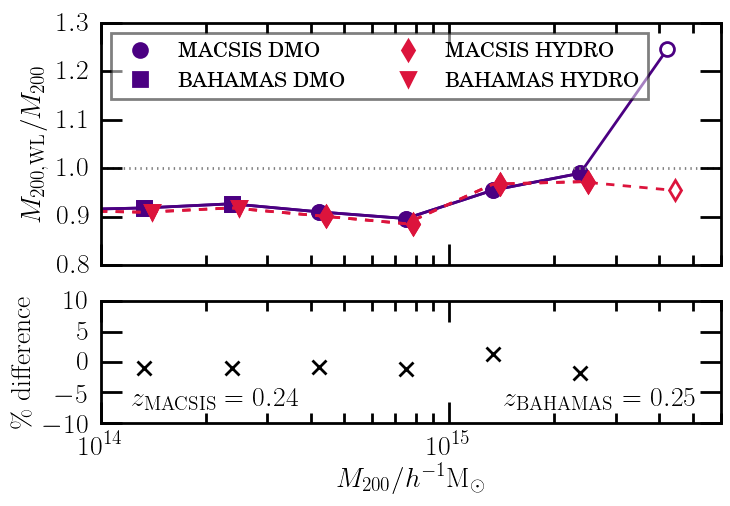}
\includegraphics[width=\linewidth]{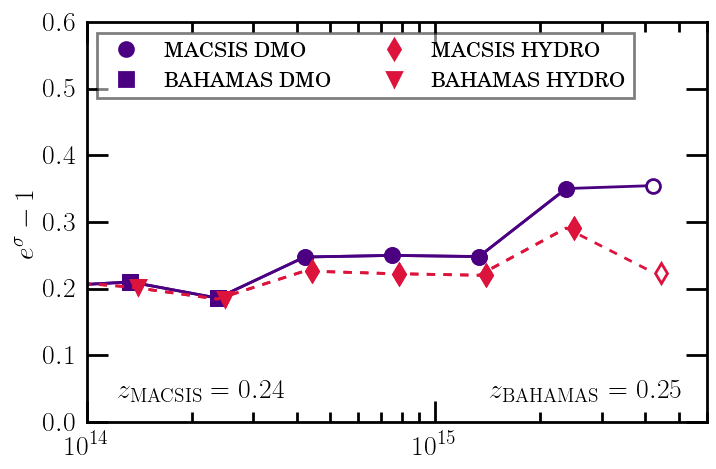}
\caption{The top panel shows the median ratio of the weak lensing mass to the true mass in mass bins obtained by fitting NFW profiles to reduced tangential shear maps for the DMO (purple solid line) and HYDRO (pink dashed line) simulations as a function of mass. Circles and diamonds (squares and triangles) are used to indicate results from the MACSIS (BAHAMAS) simulations. Markers are offset horizontally for clarity. The unfilled markers represent bins containing fewer than 10 projections. The percentage difference between the DMO and HYDRO result is shown in the middle panel. The bottom panel shows the scatter in the bias as a function of mass, where $\sigma$ is the standard deviation.}
\label{fig:wl-mbias-baryons}
\end{figure}

\begin{figure*}
\subfloat{\includegraphics[width=0.5\linewidth]{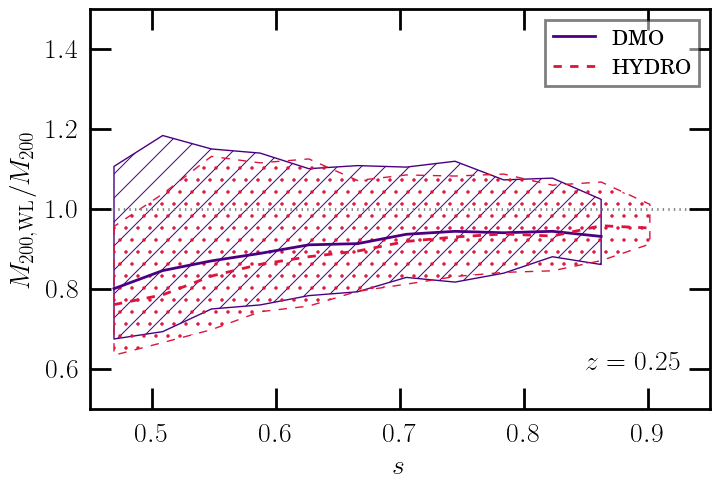}}
\subfloat{\includegraphics[width=0.5\linewidth]{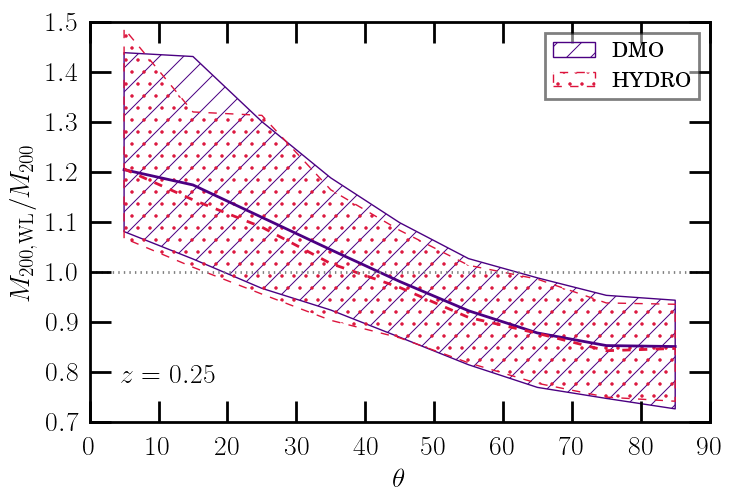}}\\
\subfloat{\includegraphics[width=0.5\linewidth]{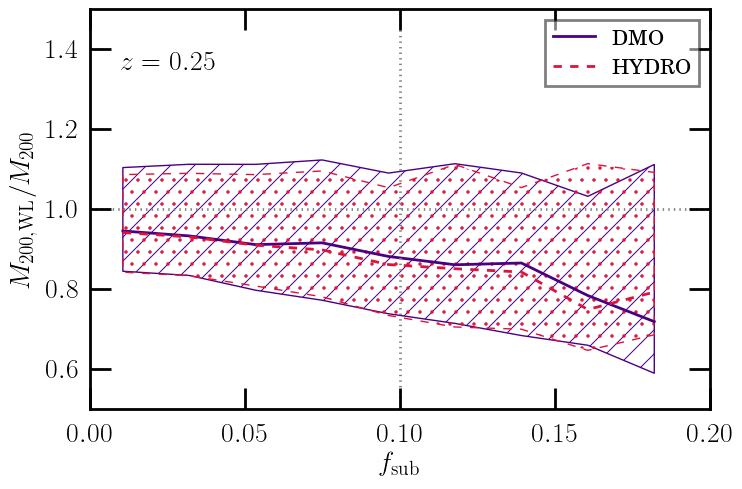}}
\subfloat{\includegraphics[width=0.5\linewidth]{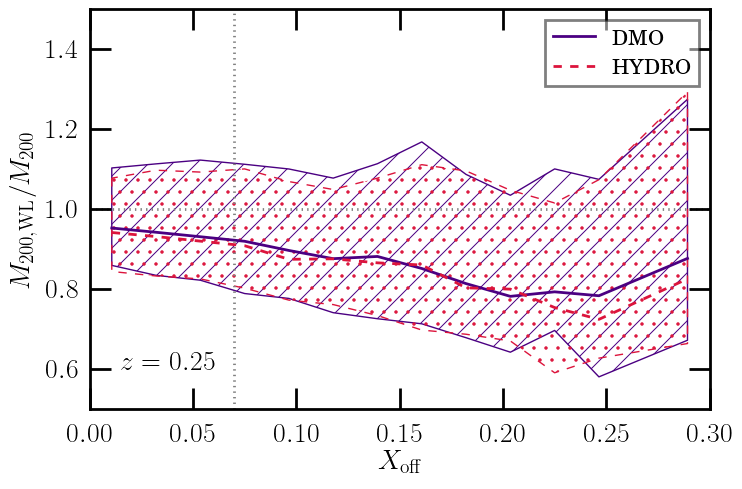}}

\caption{The dependence of the ratio of the weak lensing mass to the true mass, $M_\mathrm{200,WL}/M_\mathrm{200}$, on cluster sphericity, $s$, alignment, $\theta$, substructure fraction, $f_\mathrm{sub}$, and centre of mass offset, $X_\mathrm{off}$, at $z=0.25$ for both the DMO (purple, diagonal hatching) and HYDRO (pink, dotted hatching) simulations. The solid lines indicate median values and the hatching represents the $1\sigma$ percentiles. Vertical dashed lines in the bottom panels indicate thresholds used to define relaxed clusters. For alignment, $\theta=90^\circ$ indicates a cluster with its principal axis perpendicular to the line of sight and $\theta=0^\circ$ indicates a cluster with its principal axis parallel to the line of sight.}
\label{fig:wl-mbias-correlations}
\end{figure*}

Fig.~\ref{fig:wl-shearprof-baryons} shows the median fractional difference between the reduced shear profiles of matched clusters in the MACSIS DMO and HYDRO simulations. Since the clusters considered in this figure all have masses $M_{200}{\geq}10^{14.5}\,h^{-1}\mathrm{M}_\odot$, our results show the impact of baryons on the shear profiles of clusters in this mass range. The shear is $12{-}15$\% larger in the central regions ($r{<}150\,h^{-1}\,\mathrm{kpc}$) of clusters in HYDRO simulations as compared to DMO simulations. This is not caused by the projection of the mass distribution, since the fractional difference in the convergence profile (bottom panel of Fig.~\ref{fig:wl-shearprof-baryons}) does not show this increase in the central region. Instead, this reflects the sensitivity of the shear to the central cluster region. This is highlighted in Equation~\ref{eq:axisymmetric-shear}, which gives the shear profile for an axisymmetric halo. Since the clusters are more centrally concentrated in the HYDRO simulations, this increases in $\bar{\Sigma}({<}r)$ in the HYDRO simulations. At $r{>}0.5\,h^{-1}\mathrm{Mpc}$, clusters in the HYDRO simulations are less dense than their DMO counterparts (see the bottom panel of Fig.~\ref{fig:wl-shearprof-baryons}), which means that $\bar{\Sigma}(<r)$ then tends to the DMO value as $r$ increases. Even outside the central region, the shear profile is sensitive to the behaviour at the centre, which makes it a more sensitive probe of baryonic effects than the convergence. 

At larger radii ($r>1\,h^{-1}\mathrm{M}_\odot$), the shear profiles from HYDRO and DMO simulations agree to within 5\%. There is significant scatter around the median values, with some clusters showing no difference between HYDRO and DMO clusters in the inner regions.

The top panel of Fig.~\ref{fig:einasto-nfw-2d} shows that when fitting the reduced tangential shear profiles of clusters, the NFW model provides an adequate fit, with a median residual $g_\mathrm{rms}=0.0453^{+0.0004}_{-0.0007}$. However, the Einasto model gives a markedly better fit, with median $g_\mathrm{rms}=0.0365^{+0.0004}_{-0.0003}$. This improved fit results in a slightly better mass estimate, as is shown in the bottom panel of Fig.~\ref{fig:einasto-nfw-2d}. Considering all BAHAMAS and MACSIS clusters at $z=0.25$, the median ratio of the estimated mass to the true mass for the Einasto model is $M_\mathrm{200,WL}/M_\mathrm{200}=0.936^{+0.003}_{-0.002}$, as compared to $M_\mathrm{200,WL}/M_\mathrm{200}= 0.911^{+0.002}_{-0.003}$ for the NFW model. However, given the difficulty in obtaining observational data of sufficient quality to constrain a three parameter fit, the minor decrease in the bias when using the Einasto model is hard to justify. 

Despite the impact of baryons on the central regions of galaxy clusters, including baryons has only a marginal impact on the weak lensing mass reconstruction at $R_{200}$. This is illustrated in Fig.~\ref{fig:wl-mbias-baryons}, in which the top panel shows the median ratio of the weak lensing mass to the true mass, $M_\mathrm{200,WL}/M_\mathrm{200}$, as a function of true mass. The results from the DMO and HYDRO simulations agree to within 5\% in all but the last mass bin, which contains only 9 projections (3 clusters). This figure shows the masses inferred from fitting with an NFW profile. Fitting with an Einasto model still gives median ratios that are consistent between the HYDRO and DMO simulations. Consistent with existing work~\citep{Oguri2011,Becker2011,Bahe2012}, the bias ($b_\mathrm{WL}=1-M_\mathrm{200,WL}/M_\mathrm{200}$) decreases with increasing mass for $M_{200}{\geq}10^{15}h^{-1}\mathrm{M}_\odot$ when using either the NFW or Einasto models. 

We find a larger bias in our weak lensing mass estimates (in both DMO and HYDRO) compared with that found in~\cite{Becker2011} and~\cite{Bahe2012}, however our results are broadly consistent with~\cite{Oguri2011}. This discrepancy is more pronounced at masses $M_{200}{\approx}10^{14}\,h^{-1}\mathrm{M}_\odot$, where~\cite{Bahe2012} find a bias of $b_\mathrm{WL}{\sim}5\%$ compared to our $b_\mathrm{WL}{\sim}9\%$. There are numerous differences between their analysis and that presented here: notably the method for fitting NFW profiles and the resolution of the underlying simulations. We fit NFW profiles to azimuthally averaged shear profiles, whereas~\cite{Bahe2012} fit directly to background ellipticities. The simulations used here have a dark matter particle mass of $5.2{\times}10^9\,h^{-1}\mathrm{M}_\odot$, which is roughly an order of magnitude larger than the particle mass in the Millennium Simulations used by~\cite{Bahe2012}. We also ignore the effect of shape noise, which has been shown to reduce the bias in weak lensing mass estimates by ${\sim}2\%$~\citep{Bahe2012}. These differences are expected to affect both the HYDRO and DMO simulations equally, which means that they do not affect our main conclusion: including baryons has no significant effect upon the bias in weak lensing mass estimates.  Note however, that baryons do change the true mass, $M_{200}$, and that studies comparing with DMO mass predictions will still obtain biased results. 

Since the bias is lognormally distributed, the scatter around these points is $e^\sigma-1$, where $\sigma$ is the standard deviation of the distribution of the bias in mass bins. This is presented in the bottom panel of the same figure. At masses greater than $4\,{\times}\,10^{14}h^{-1}\mathrm{M}_\odot$, a difference in the scatter in DMO and HYDRO simulations emerges. Whilst the scatter is dominated by the alignment of the cluster with the line of sight~\citep[in agreement with][]{Bahe2012}, this difference is likely due to substructures. For $M_{200}{\geq}10^{14}h^{-1}\mathrm{M}_\odot$ at $z=0.25$, HYDRO clusters have smaller values of $f_\mathrm{sub}$ on average than their DMO counterparts. 

\subsection{The correlation of weak lensing mass bias with cluster parameters}

Fig.~\ref{fig:wl-mbias-correlations} shows the dependence of the bias in weak lensing mass estimates on sphericity, alignment, the substructure fraction and the centre of mass offset. Since $M_\mathrm{200,WL}/M_\mathrm{200}$ tends to unity with increasing mass, and sphericity decreases with increasing mass, the increase of $M_\mathrm{200,WL}/M_\mathrm{200}$ with increasing sphericity is not simply a by-product of the mass dependence of the sphericity. The top left image in Fig.~\ref{fig:wl-mbias-correlations} shows that there is also  an offset between the DMO and HYDRO simulations when plotting against sphericity. This offset is also present in the top right image in the same figure, which shows the mass bias as a function of the alignment of the cluster, $\theta$, where $\theta$ is the angle between the longest semi-principal axis of the cluster and the line of sight. For $\theta{<}43^\circ$ ($\theta{>}43^\circ$) the masses of clusters that are elongated along the line of sight are on average overestimated (underestimated). 

The bottom panels of Fig.~\ref{fig:wl-mbias-correlations} show the dependence of the weak lensing mass bias on two parameters typically used to characterise the dynamical state of clusters in simulations: the substructure fraction and the centre of mass offset. The more substructure in a cluster, the greater the bias in its weak lensing mass. A similar trend is present for $X_\mathrm{off}$, with the masses of clusters exhibiting a large centre of mass offset being underestimated by around $20\%$. There appears to be an upturn at $X_\mathrm{off}{>}0.25$ in both the DMO and HYDRO simulations, with the last bin consisting of 12 (15) projections in the DMO (HYDRO) simulations. From a visual inspection of the projected mass distributions of these clusters, this appears to be a consequence of substructures appearing to be in the central cluster regions due to projection effects. Whilst this only occurs in one projection per cluster, in these cases it has led to such a large over estimate of the mass that the median bias is a higher than at smaller values of $X_\mathrm{off}$.

\section{Hydrostatic Bias}
Cluster weak lensing is only one method for estimating cluster masses. Another widely used approach is to calculate masses from the gas temperature and density profiles derived from X-ray observations. This approach assumes that clusters are in hydrostatic equilibrium, which may lead to a bias in the resulting masses. This hydrostatic mass bias has been studied thoroughly in the literature, yet a consistent narrative is yet to emerge. The bias is typically parametrised in terms of $b_\mathrm{X}=1-M_\mathrm{X}/M_\mathrm{WL}$, where $M_\mathrm{X}$ is the hydrostatic mass obtained from a cluster's X-ray emission.

For observations, ~\cite{Applegate2014},~\cite{Israel2014} and~\cite{Smith2016} find a bias of only $b_\mathrm{X}=0.04,0.08$ and $0.05$ respectively. Yet this is at odds with results from~\cite{vonderLinden2014},~\cite{Hoekstra2015} and ~\cite{Simet2015}, who find $b_\mathrm{X}=0.30,0.24$ and $0.22$ respectively. The hydrostatic bias has been shown to be larger at larger radii~\citep{Zhang2010}, which may go some way to explaining the result from~\cite{Applegate2014}, where it is measured at $R_{2500}$. However, the results from~\cite{Israel2014} and~\cite{Smith2016} are measured at $R_{500}$. 
It should also be noted that these studies use different X-ray datasets; for example~\citep{Applegate2014} uses a sample of 12 relaxed clusters, whereas~\citep{vonderLinden2014} use a subset of clusters from the Planck catalogue~\citep{Planckclusters2014} that are also present in the Weighing the Giants sample~\citep{vonderLindenwtg2014}.Since these studies span overlapping mass ranges, mass can also not account for the disparity. ~\cite{Smith2016} attribute the difference to redshift, suggesting that the bias at $z{>}0.3$ may be larger as a consequence of observational systematics. 

Yet this would not explain the contrast between the results of ~\cite{Applegate2014},~\cite{Israel2014} and~\cite{Smith2016} and the results from numerical simulations. Numerical simulations have typically found a bias of ${\sim}20\%$ for groups and clusters~\citep{Nagai2007,Kay2012,Rasia2012,LeBrun2014}, with some dependence on the implementation of baryonic physics~\citep{Kay2012,LeBrun2014}. Since simulating massive clusters with hydrodynamics (including realistic feedback) is computationally expensive, most of these works have not had sufficient numbers of high mass clusters to characterise the mass dependence of the hydrostatic bias. Combining the MACSIS and BAHAMAS samples gives a large sample of clusters spanning $10^{14}{\leq}M_{500}/h^{-1}\mathrm{M}_\odot{\leq}10^{15}$. 

\begin{figure}
\centering
\includegraphics[width=\linewidth]{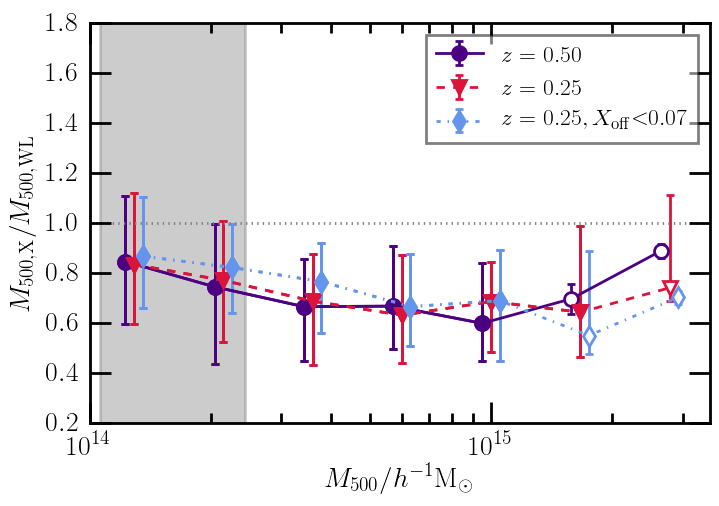}
\includegraphics[width=\linewidth]{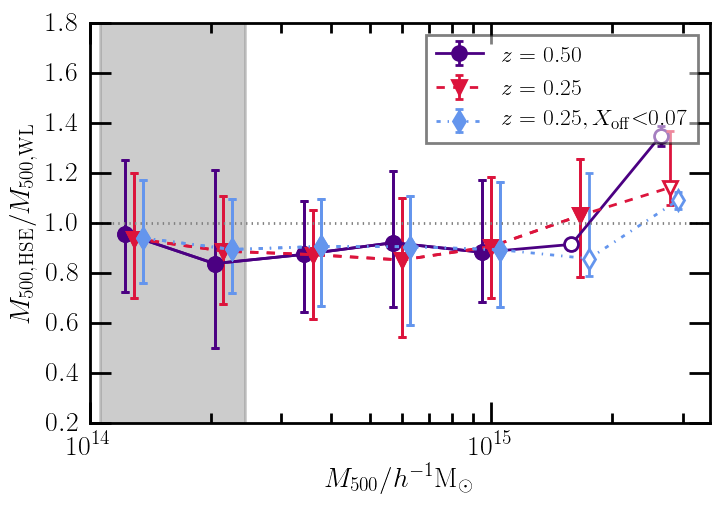}

\caption{The dependence of the ratio of the X-ray hydrostatic mass to the mass inferred from weak lensing, $M_\mathrm{500,X}/M_\mathrm{500,WL}$, on the true mass of the cluster at $z=0.25,0.5$.  The top panel shows hydrostatic masses derived using spectroscopic temperatures and densities. The bottom panel shows hydrostatic masses derived using true temperatures and densities. Weak lensing masses and hydrostatic masses are calculated independently.  The shaded grey region shows the mass range where the MACSIS and BAHAMAS simulations overlap. Excluding the highest mass bin (which contains fewer than 10 clusters), the ratio decreases as a function of mass when using spectroscopic values. The mass bias is independent of redshift. Selecting relaxed clusters (clusters with $X_\mathrm{off}{<}0.07$) has no significant effect on the bias. Weak lensing masses were obtained by taking one random projection of each cluster. The markers are offset horizontally for clarity.}
\label{fig:hydrostatic-bias-mass}
\end{figure}

To obtain hydrostatic masses for clusters in the BAHAMAS and MACSIS simulations, the X-ray spectra of all particles within $5R_{200}$ are calculated and then binned into 25 radial bins spaced logarithmically between $0.03R_{200}$ and $5R_{200}$. As described in~\cite{LeBrun2014}, gas particles with temperature $k_BT{<}10^{5.2}$\,keV and number density $n{\geq}0.1\,\mathrm{cm}^{-3}$ are excluded. Modifying the threshold of these cuts by up to an order of magnitude or excluding bound substructures has no meaningful effect on our results. In each radial bin, the emission spectrum is fitted with a single temperature APEC model, giving a temperature, $T_X$, density, $\rho_X$, and metallicity. We find assuming a fixed metallicity leads to a larger bias in the measured temperatures, so we fit for the temperature, density and metallicity simultaneously. We fit the spectra in the range $0.05-10.0$\,keV and extending this range to $0.05-20.0$\,keV does not affect our results. We fit the density and temperature profiles using the functional forms in~\cite{Vikhlinin2006}. Under the assumption of hydrostatic equilibrium, these are used to infer a mass profile and thus a mass.

\begin{figure}
\centering
\includegraphics[width=\linewidth]{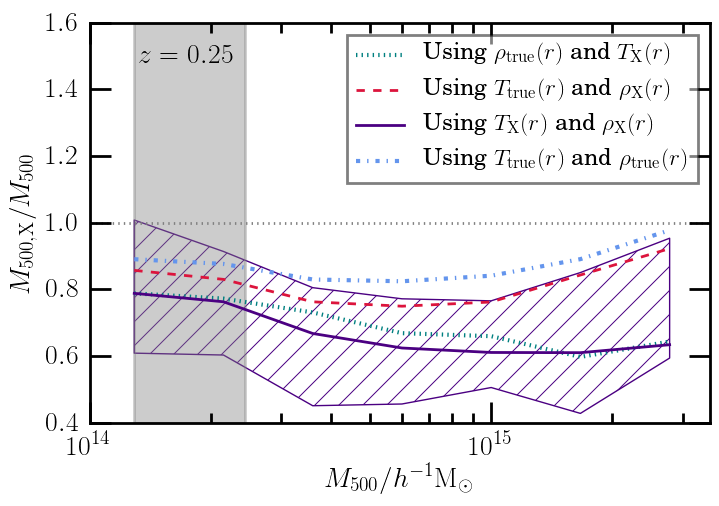}
\caption{The ratio of X-ray mass to true mass when using the mass weighted temperature $T_\mathrm{true}(r)$ profile and the true gas density profile $\rho_\mathrm{true}(r)$ (line blue dot-dashed line) as compared to the bias when using the X-ray temperature profile, $T_\mathrm{X}(r)$ and the X-ray density profile $\rho_\mathrm{true}(r)$ (solid purple line). The hatched region shows the scatter in the bias obtained when using X-ray observables. The shaded grey region shows the mass range where the MACSIS and BAHAMAS simulations overlap. Masses obtained from X-ray observables are consistently lower, with the X-ray temperature providing the largest contribution (some percentage here) to the mass bias.}
\label{fig:hydrostatic-bias-temps-density}
\end{figure}

\begin{figure*}
\subfloat{\includegraphics[width=0.5\linewidth]{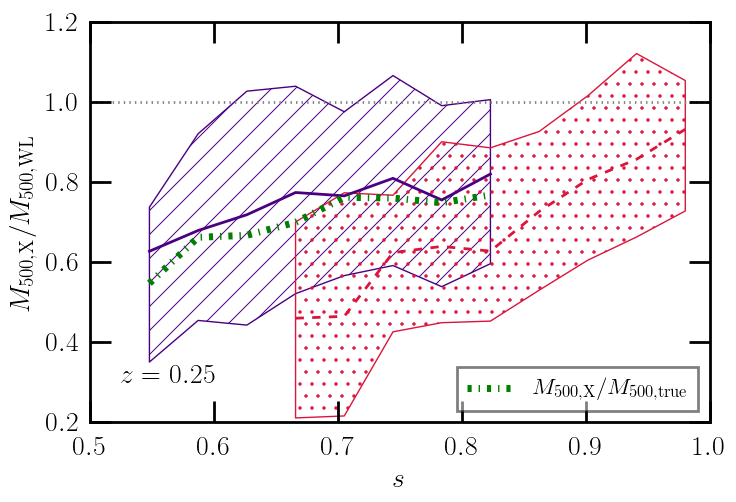}}
\subfloat{\includegraphics[width=0.5\linewidth]{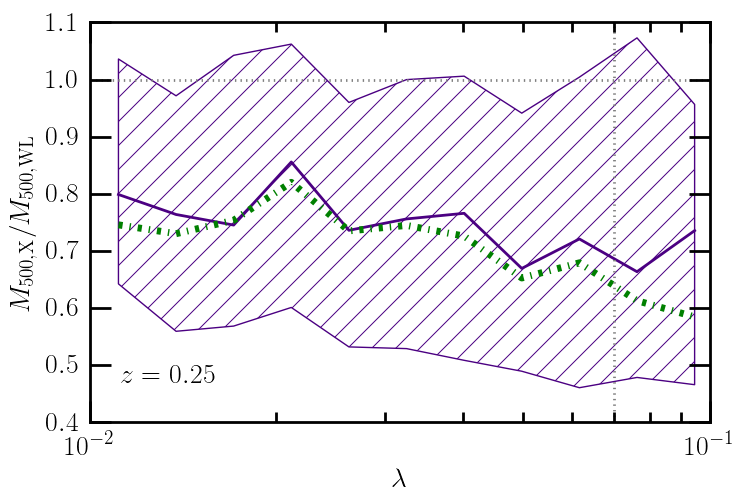}} \\[2pt]
\subfloat{\includegraphics[width=0.5\linewidth]{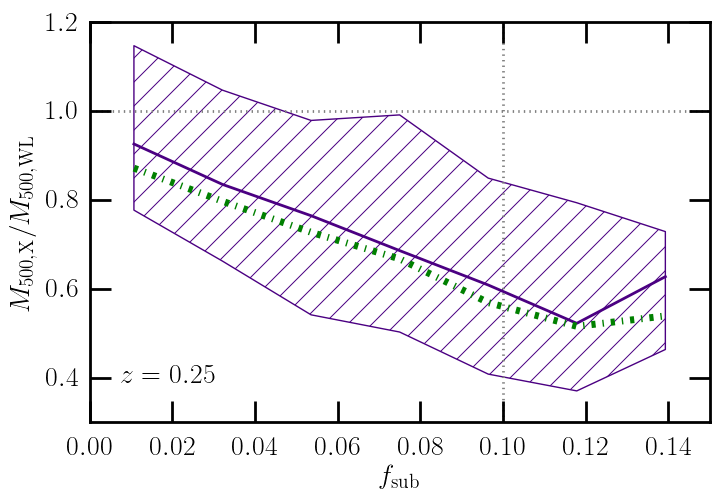}}
\subfloat{\includegraphics[width=0.5\linewidth]{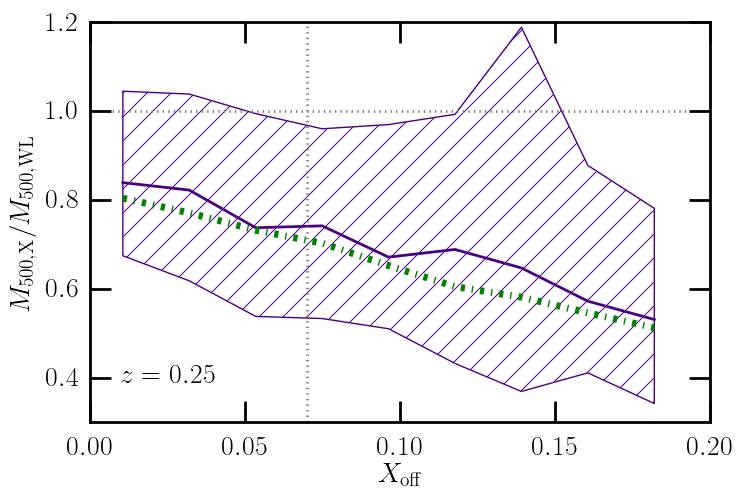}}
\caption{The dependence of ratio of the X-ray hydrostatic mass to the mass inferred from weak lensing, $M_\mathrm{500,X}/M_\mathrm{500,WL}$, on cluster sphericity, $s$, spin, $\lambda$, substructure fraction, $f_\mathrm{sub}$, and $X_\mathrm{off}$ at $z=0.25$ for the HYDRO simulations.  The green dot-dashed line shows the dependence of the ratio of the X-ray hydrostatic mass to the true mass, $M_\mathrm{200,X}/M_\mathrm{200,true}$, on the same parameters. The thick solid lines indicate median values in mass bins and the hatched regions indicate $1\sigma$ percentiles. The hydrostatic bias is measured for the total mass distribution and the purple solid lines indicate parameters measured for the total mass distribution. The pink dashed line in the top left panel indicates the sphericity of the gas distribution only. The sphericity and spin parameter are measured within $R{200}$. The substructure fraction and centre of mass offset are measured within the virial radius. The hydrostatic mass bias shows a stronger dependence on the sphericity of the gas than on the sphericity of the total matter distribution. Weak lensing masses and hydrostatic masses are calculated independently. Vertical dashed lines indicate thresholds used to define relaxed clusters. }
\label{fig:hsebias-correlations}
\end{figure*}
Fig.~\ref{fig:hydrostatic-bias-mass} shows the ratio of the X-ray hydrostatic mass to the mass inferred from weak lensing, $1{-}b=M_\mathrm{500,X}$/$M_\mathrm{500,WL}$, as a function of true mass, $M_{500}$. Weak lensing masses and hydrostatic masses are calculated independently. Weak lensing masses are calculated using the approach described in Section~\ref{sec:wl-profiles}, replacing 200 for 500 in Equation~\eqref{eq:wl-r200} and Equation~\eqref{eq:wl-m200}. Our findings for $M_\mathrm{200,WL}/M_{200}$ are consistent with those for $M_{500,WL}/M_{500}$.  

In the top panel of Fig.~\ref{fig:hydrostatic-bias-mass}, which shows hydrostatic masses obtained from spectroscopic temperature and density profiles, we see that the hydrostatic mass is consistently smaller than the weak lensing mass for clusters of all masses. The bias increases from $b_\mathrm{X}=0.2$ to $0.35$ as the mass increases from $10^{14}\,h^{-1}\mathrm{M}_\odot$ to $10^{15}\,h^{-1}\mathrm{M}_\odot$. There appears to be an upturn a $M_{500}{\geq}10^{15}\,h^{-1}\mathrm{M}_\odot$, however since the highest mass bin contains only nine projections (three clusters), this may be a consequence of limited statistics. There is no evidence for any redshift dependence. As the light blue dot-dashed line shows, selecting relaxed clusters does not reduce the bias. We have confirmed that the spectra and functional forms for the density and temperature profiles are well fitted. 

Whilst the bias in X-ray mass measurements is referred to as the hydrostatic bias, it does not only reflect the bias due to the assumption that the cluster is hydrostatic. As the bottom panel of Fig.~\ref{fig:hydrostatic-bias-mass} shows, if a hydrostatic mass is calculated using the true density and temperature profiles, $\rho_\mathrm{true}(r)$ and $T_\mathrm{true}(r)$, then the bias is significantly reduced to $b_\mathrm{X}=0.04{-}0.14$ for all bins containing at least 10 clusters. This difference is independent of redshift and is still present when considering only relaxed haloes.

Figure~\ref{fig:hydrostatic-bias-temps-density} shows that the bias in the X-ray temperature profile is the dominant contribution to the mass bias, since using the density profile derived from X-ray observations has very little effect. This is consistent with the work of~\cite{Rasia2012} and~\cite{Biffi2016}.~\cite{Rasia2012} noted that the temperature inhomogeneities in the intracluster medium lead to lower X-ray temperatures, which in turn leads to smaller X-ray masses. They suggested that including feedback from AGN would reduce this effect, however it is clearly still present in both these simulations and those studied in~\cite{Biffi2016}. We have extensively tested our analysis pipeline and find that the lower X-ray temperatures are a consequence of cooler gas (with $T{\sim}1{-}3$keV) emitting X-rays in the cluster outskirts. Since excluding gas in substructures does not affect our results, we can infer that this gas is not bound in substructures. Further work is needed to investigate the origin of this gas; it may be gas that is stripped from infalling substructures or it may be cool accreted gas. Since these simulations use a traditional SPH scheme, which has been shown to lead to a lack of mixing in the simulations~\citep{Sembolini2016a}, it is possible that the presence of this cool gas may be an unphysical artefact that arises from a lack of mixing in the intracluster medium. However,~\cite{Sembolini2016b} found that the differences in cluster cores due to the hydrodynamics solver are overwhelmed by differences due to the inclusion of AGN feedback and differences in its implementation. Furthermore, it is not yet clear what amount of mixing is realistic, since it may depend on other effects not considered here (e.g. the magnetic field structure of the intracluster medium). 

If we consider the hydrostatic mass calculated using $\rho_\mathrm{true}(r)$ and $T_\mathrm{true}(r)$ (rather than the gas density and temperature profiles as obtained from mock X-ray observations) to be representative of the hydrostatic bias, then we find $M_\mathrm{500,X}$/$M_\mathrm{500,true}{\approx}0.8{-}0.9$ for clusters in the mass range $10^{14}{\leq}M_{500}/h^{-1}\mathrm{M}_\odot{\leq}3{\times}10^{15}$.

Finally, in Fig.~\ref{fig:hsebias-correlations} we show the dependence of the ratio of the X-ray hydrostatic mass (calculated using spectral temperatures and densities) to the mass inferred from weak lensing as a function of sphericity, spin, substructure fraction and centre of mass offset. The bias shows a strong dependence on sphericity, with a smaller bias for more spherical clusters. For sphericity, the purple line indicates the hydrostatic bias against the sphericity of the total mass distribution, whereas the pink line is the bias against the sphericity of the gas. The hydrostatic bias exhibits a tighter correlation and a stronger mass dependence with gas sphericity than with the total matter sphericity. This is not driven by the weak lensing mass estimate, since the same trend is present in $M_\mathrm{500,X}/M_\mathrm{true}$ (the green dot-dashed line). Instead this justifies the morphological selections used in X-ray observations~\citep{Postman2012}, since it suggests that clusters that are more spherical will exhibit a smaller hydrostatic mass bias. 

Despite the frequent use of spin as an indicator of the dynamical state of a cluster in simulations~\citep[e.g.][]{Klypin2016}, we find that the hydrostatic bias is largely independent of spin, regardless of the approach used to calculate the hydrostatic mass. Rather, it shows a strong dependence on the substructure fraction, with higher values of $f_\mathrm{sub}$ yield a larger bias. This dependence persists even when cool dense clumps are not removed from the cluster prior to calculating hydrostatic masses. Since we find that X-ray masses are less biased for clusters that are more spherical and contain less substructure, this suggests that the bias is lower for older clusters, which have accreted less material recently. 

If the substructure fraction reflects the appearance of multiple peaks in the X-ray emission, then this motivates the X-ray selection techniques based on a visual identification of substructure~\cite[e.g.][]{Nurgaliev2013,Rasia2013b}. We defer a more detailed study of the former assumption to later work. Finally, the hydrostatic bias also correlates with the centre of mass offset. Yet, the large scatter in these relations demonstrates that even the hydrostatic masses of clusters with small values of $f_\mathrm{sub}$ and $X_\mathrm{off}$ can be biased low by up to 20\%. 

\section{Summary}
In this study we have used the MACSIS and BAHAMAS simulations presented in~\cite{Barnes2016} and~\cite{McCarthy2016} to create a combined sample of more than 3,500 clusters with $M_{200}{\geq}5{\times}10^{13}\,h^{-1}\mathrm{M}_\odot$, simulated with realistic baryonic physics. These simulations have been shown to reproduce the observed scalings of gas mass, integrated Sunyaev-Z'eldovich signal and X-ray luminosity with mass, as well as the observed hot gas radial profiles of clusters at $z=0$~\citep{McCarthy2016,Barnes2016}.

We focus our study on three key areas: the properties of high mass clusters and the impact of baryons upon them, the influence of baryonic effects upon weak lensing mass estimates, and the mass dependence of the hydrostatic bias in high mass clusters.

Since the MACSIS simulations consist of matched HYDRO and DMO zoom simulations, we are able to directly compare clusters simulated with and without baryonic effects. We also investigated the redshift dependence of our results. Our main results are as follows:
\begin{itemize}
\item The distributions of spins in the HYDRO and DMO simulations are consistent with each other and are well fitted by a lognormal distribution. The dark matter component has a slightly larger spin in the HYDRO simulations than in the DMO simulations, which is associated with a transfer in angular momentum from the baryonic component to the dark matter. Spin declines weakly with mass at all redshifts considered here (Fig.~\ref{fig:spin-shape-mass}). The slope is consistent between the HYDRO and DMO simulations and is unchanged for a relaxed subsample. The mean spin of relaxed haloes is 15\% smaller than for the entire cluster sample. 
\item Clusters in the HYDRO simulations are more spherical, with larger values of $s$ and $e$ on average. The sphericity-mass relation is steeper in the DMO simulations, but the elongation-mass relation is consistent between the DMO and HYDRO simulations. A larger mass range is required to constrain the effect of baryons on this slope. Selecting only relaxed haloes does not affect the slope of either the sphericity-mass or elongation-mass relation.
\item By matching MACSIS clusters in the DMO and HYDRO simulations we demonstrated that clusters in the HYDRO simulations are more concentrated in the central regions (Fig.~\ref{fig:macsis-dmo-hydro-profdiff}). This is partly due to the condensation of baryons in the cluster centre and also a consequence of the contraction of the dark matter halo in the presence of baryons. The dark matter density profiles of MACSIS clusters at $z=0$ in the HYDRO simulations are more dense at $r{<}0.6R_{200}$ than in the DMO simulations. At $0.6{<}r/R_{200}{<}3$ the dark matter density profile is less dense in the HYDRO simulations than in the DMO simulations. At the high-mass end of the concentration mass relation ($M_{200}>10^{15}h^{-1}\mathrm{M}_\odot$) this manifests itself as an increase in concentrations in the HYDRO simulations (Fig.~\ref{fig:c200-nfw-mass}). Since the concentration-mass relation is flatter in the HYDRO simulations, clusters with masses $M_{200}{\approx}10^{15}h^{-1}\mathrm{M}_\odot$ have larger concentrations than clusters in the DMO simulations.
\item The density profiles of clusters considered here are better fit by the Einasto profile than the NFW profile. This leads to a smaller bias in the masses calculated from fits to the spherically averaged density profile, with the NFW model underpredicting masses by 22\% on average and the Einasto model underpredicting masses by 8\% (Fig.~\ref{fig:einasto-nfw-3d}). Whilst cluster shear profiles are better fit by the Einasto model rather than the NFW model, this only results in a $2{-}3$\% improvement in weak lensing cluster mass estimates in the HYDRO simulations, despite the cost of adding an additional degree of freedom (Fig~\ref{fig:einasto-nfw-2d}).
\item Baryons have a more significant effect on the shear profiles of clusters than on their convergence profiles. The shear profiles of HYDRO clusters are up to 15\% larger than clusters in the DMO simulations at $r{<}0.5h^{-1}\mathrm{M}_\odot$, as a consequence of the sensitivity of the shear to the central cluster region (Fig.~\ref{fig:wl-shearprof-baryons}). Despite this, the weak lensing mass bias is consistent between the DMO and HYDRO simulations, with both data sets showing that weak lensing underestimates cluster masses by ${\approx}10\%$ for clusters with $M_{200}{\leq}10^{15}h^{-1}\mathrm{M}_\odot$ and that this bias tends to zero at higher masses (Fig.~\ref{fig:wl-mbias-baryons}). 
\item The hydrostatic mass bias, $1-b=M_\mathrm{500,X-ray}/M_\mathrm{500,WL}$ declines from $0.8$ to $0.6$ for clusters with masses increasing from $M_{500}=10^{14}h^{-1}\mathrm{M}_\odot$ to $M_{500}=10^{15}h^{-1}\mathrm{M}_\odot$ when using X-ray hydrostatic masses calculated from spectroscopic temperature and density profiles (Fig.~\ref{fig:hydrostatic-bias-mass}). The X-ray and weak lensing masses are measured independently. We find no evidence for any redshift dependence. The mass dependence is mostly due to the spectroscopic temperature measurements (Fig~\ref{fig:hydrostatic-bias-temps-density}) that are biased low by the presence of cooler, X-ray emitting gas in the cluster outskirts. Using the true temperature and density profiles gives $b{\approx}0.04{-}0.14$ at the masses considered here, with no clear mass dependence.
\item The hydrostatic bias is smaller for more spherical clusters that have a small centre of mass offset and fewer substructures, which motivates the morphological selection of clusters in X-ray surveys (Fig.~\ref{fig:hsebias-correlations}).
\end{itemize}

In conclusion, we find baryons have only a minor effect on the spins, shapes and weak lensing mass estimates of massive galaxy clusters. Baryons have a small effect on cluster density profiles at small radii, which is also apparent in their weak lensing shear profiles. When using spectroscopic temperatures and densities, the hydrostatic bias decreases as a function of mass, leading to a bias of ${\approx}40\%$ for high mass clusters. Further work is needed to clarify the cause of this large bias and to reconcile it with observational results.

\section*{Acknowledgements}
This work used the DiRAC Data Centric system at Durham University, operated by the Institute for Computational Cosmology on behalf of the STFC DiRAC HPC Facility (www.dirac.ac.uk). This equipment was funded by BIS National E-infrastructure capital grant ST/K00042X/1, STFC capital grants ST/H008519/1 and ST/K00087X/1, STFC DiRAC Operations grant ST/K003267/1 and Durham University. DiRAC is part of the National E-Infrastructure. MAH is supported by an STFC quota studentship. DJB and STK acknowledge support from STFC through grant ST/L000768/1. IGM is supported by a STFC Advanced Fellowship. The research was supported in part by the European Research Council under the European Union's Seventh Framework Programme (FP7/2007-2013) / ERC Grant agreement 278594-GasAroundGalaxies.

%Bibliography
\bibliographystyle{mnras} %Find better style file
\bibliography{references}
\label{lastpage}

\end{document}